\documentclass{emulateapj}
\usepackage[dvipsnames,usenames]{color}
\usepackage{colortbl}
\usepackage{graphicx}
\usepackage{amsmath}
\usepackage[]{natbib}
\usepackage[usenames]{color}
\usepackage{float}
\usepackage{subfigure}
\usepackage{color}
\usepackage{hyperref}
\usepackage{float}

\newcommand{\sauron}{\texttt{SAURON}}
\newcommand{\atlas}{\texttt{ATLAS$^{3D}$}}
\newcommand{\ppxf}{\texttt{pPXF}}
\newcommand{\galaxy}{\texttt{GALAXY }}

\newcommand{\refsec}[1]{Section~\ref{#1}}
\newcommand{\reffig}[1]{Fig.~\ref{#1}}

\begin{document}
\title{Kinematic properties of double-barred galaxies: simulations vs. integral-field observations}

\author{
Min Du\altaffilmark{1,2},
Victor P. Debattista\altaffilmark{3},
Juntai Shen\altaffilmark{1},
Michele Cappellari\altaffilmark{4}}

\altaffiltext{1}{Key Laboratory of Research in Galaxies and Cosmology, Shanghai Astronomical Observatory, Chinese Academy of Sciences, 80 Nandan Road, Shanghai 200030, China}
\altaffiltext{2}{Correspondence should be addressed to jshen@shao.ac.cn; vpdebattista@gmail.com}
\altaffiltext{3}{Jeremiah Horrocks Institute, University of Central Lancashire, Preston, PR1 2HE, UK}

\altaffiltext{4}{Sub-department of Astrophysics, Department of Physics, University of Oxford, Denys Wilkinson Building, Keble Road, Oxford OX1 3RH}

\begin{abstract}
  
Using high resolution $N$-body simulations, we recently reported that
a dynamically cool inner disk embedded in a hotter outer disk can
naturally generate a steady double-barred (S2B) structure.  Here we
study the kinematics of these S2B simulations, and compare them to
integral-field observations from \atlas and \sauron.
We show that S2B galaxies exhibit several distinct
kinematic features, namely: (1) significantly distorted isovelocity
contours at the transition region between the two bars, (2) peaks in
$\sigma_\mathrm{LOS}$ along the minor axis of inner bars, 
which we term ``$\sigma$-humps'', that are often accompanied by ring/spiral-like 
features of increased $\sigma_\mathrm{LOS}$, (3) $h_3-\bar{v}$ anti-correlations in 
the region of the inner bar for certain orientations, and (4) rings of
positive $h_4$ when viewed at low inclinations. The most impressive
of these features are the $\sigma$-humps; these evolve with the
inner bar, oscillating in strength just as the inner bar does
as it rotates relative to the outer bar. We show that, in
cylindrical coordinates, the inner bar has similar streaming
motions and velocity dispersion properties as normal large-scale bars,
except for $\sigma_z$, which exhibits peaks on the minor axis,
i.e., humps. These $\sigma_z$ humps are responsible for
producing the $\sigma$-humps. For three 
well-resolved early-type S2Bs (\object{NGC 2859}, \object{NGC 2950}, 
and \object{NGC 3941}) and a potential S2B candidate (\object{NGC 3384}), the
S2B model qualitatively matches the integral-field data well, 
including the ``$\sigma$-hollows'' previously identified.
We also discuss the kinematic effect of a nuclear disk in S2Bs.

\end{abstract}


\section{Introduction}

Optical and infrared observations have shown that $\sim1/3$ of
early-type barred galaxies host a misaligned inner bar (also
termed ``secondary bar'') \citep{erw_spa_02,lai_etal_02, erw_04}. S2B
structures are also seen in later Hubble types, although we still lack
systematic statistics because of the stronger dust extinction in their
central regions \citep{erw_05}. Numerical simulations are powerful
tools for studying the formation and evolution of such multi-bar
structures. Previous $N$-body+hydrodynamical simulations suggested
that gas dissipation plays a vital role in inducing and maintaining an
inner bar \citep[e.g.][]{fri_mar_93, shl_hel_02, eng_shl_04}. 
However, the observation of galaxies without a large
amount of gas \citep{pet_wil_04} indicated gas might not be the key
ingredient to maintaining, or even forming, S2Bs. Increasingly,
$N$-body simulations have successfully formed S2Bs without the
requirement of gas \citep{rau_sal_99, rau_etal_02, deb_she_07,
hel_etal_07, sah_mac_13, du_etal_15}. Nevertheless, the essential initial
conditions by which S2Bs form is still unclear. In \citet{du_etal_15},
we explored a large parameter space of the mass, dynamical temperature
(Toomre-$Q$), and thickness of the stellar disk in isolated pure-disk
3-D $N$-body simulations. Our simulations suggested that a dynamically
cool inner disk can naturally trigger small-scale bar instabilities
leading to S2Bs, without the need for gas. This result is also
consistent with the result of \citet{woz_15}, who successfully formed
long-lived S2Bs with $N$-body+hydrodynamical simulations in which a
nuclear disk forming from accumulated gas followed by star formation
which plays an important role in generating the inner bar.
This scenario is also consistent with the recent observation of 
\object{NGC 6949} that the size of the star burst nuclear molecular 
disk matches well with the size of the inner bar \citep{rom_fat_15}.

Observations \citep{but_cro_93,fri_mar_93, cor_etal_03} suggest that
the two bars in an S2B rotate independently, which is also found in
numerical simulations \citep[e.g.][]{deb_she_07,
she_deb_09, sah_mac_13, woz_15, du_etal_15}. Instead of being rigid
bodies, the amplitudes and pattern speeds oscillate as the two bars rotate
through each other \citep{deb_she_07, du_etal_15}, which is consistent
with loop-orbit predictions of \citet{mac_ath_07} \citep[see
also][]{mac_spa_97, mac_spa_00, mac_ath_08, mac_sma_10}. Such
dynamically decoupled inner bars in S2Bs have been hypothesized to
be a mechanism for driving gas past the inner Lindblad resonance (ILR)
of outer bars to feed supermassive black holes that power active
galactic nuclei \citep{shl_etal_89,shl_etal_90}.

Two-dimensional integral-field unit (IFU) spectroscopy provides a very
powerful method for studying bars from a kinematic point of
view. Several kinematic signatures of bars have been predicted
and observed. Many theoretical analyses \citep[e.g.][]{mil_smi_79,
vau_dej_97, bur_ath_05} have shown that bars twist the mean
velocity ($\bar{v}$) fields because of significant radial streaming motions, 
thus making the kinematic major axis misaligned with the photometric 
major axis of the whole disk. For both stars and gas, the kinematic major 
axis generally turns toward the opposite direction with respect to the 
major axis of bars. IFU observations of the early-type galaxies 
have shown that barred galaxies are more likely 
to generate larger kinematic misalignments than unbarred galaxies  
\citep{cap_etal_07, kra_etal_11}. 
The central elliptical velocity dispersion ($\sigma$) peak
should be aligned with the orientation of the large-scale bar
\citep{mil_smi_79, vau_dej_97}. Over the extent of the bar, the third
order Gauss-Hermite moment ($h_3$) is correlated with $\bar{v}$ in
edge-on views \citep{bur_ath_05}. In face-on views, a minimum in $h_4$
is present when a boxy/peanut (B/P) bulge exists \citep{deb_etal_05, mendez_etal_14}.

We know little about the kinematic properties of S2Bs. The misalignment 
between the kinematic major axis and the photometric major axis 
has also been expected to be observed in $\bar{v}$ fields of S2Bs
\citep{che_fur_78, moi_mus_00}. However, \citet{moi_etal_04} found the twists
due to the inner bar on the stellar velocity field are quite
small compared with the twists in gaseous kinematics, which led them
to question the existence of decoupled inner bars.
\citet{she_deb_09} showed that twists due to inner bars are
smaller than previously expected, thus the kinematics of S2Bs can
still be consistent with observations of \citet{moi_etal_04}.
\citet{de_etal_08} studied 2-D stellar velocity and velocity dispersion 
maps of four S2Bs (\object{NGC 2859}, \object{NGC 3941}, \object{NGC 4725}, and 
\object{NGC 5850}) with the \sauron \ IFU. Surprisingly, the velocity 
dispersion maps revealed two local minima, which they termed ``$\sigma$-hollows'', 
located near the ends of the inner bar in each galaxy \citep[see also][]{de_etal_12}. 
They proposed that $\sigma$-hollows occur as a result of the contrast between 
the velocity dispersion of a hotter bulge and the inner bar which is 
dominated by ordered motions and thus has a low $\sigma$. The S2B model of
\citet{she_deb_09} also exhibited a misalignment between the
inner bar and the velocity dispersion.

Self-consistent numerical models are very powerful tools for understanding the 
dynamics and kinematics of S2Bs. In \citet{du_etal_15}, we were able to form 
S2Bs from pure disks; we summarize these results in \refsec{subsection:models}. 
In this paper, we analyse the kinematics of the S2B model. We introduce the 
Voronoi binning method used in extracting the kinematics in \refsec{subsection:extkinem}. 
In \refsec{section:Atlas3D}, we show that the S2B model qualitatively matches well 
with the kinematics of S2Bs in the \atlas \ \citep{cap_etal_11} and \sauron \
\citep{ems_etal_04} surveys, especially for the $\sigma$-humps/hollows. The detailed 
kinematic analyses of the S2B model are presented in \refsec{section:kinem}. 
In \refsec{section:discu}, we discuss the kinematic effects of a nuclear disk in 
the S2Bs. Finally, our conclusions are summarized in \refsec{section:conclusion}.

\section{Method}
\label{section:method}

\subsection{Models}
\label{subsection:models}

A detailed description of the initial conditions and evolution of our
self-consistent $N$-body models has been presented in
\citet{du_etal_15}. Here we give a brief introduction to the models 
we evolved with a 3-D cylindrical polar grid code, \galaxy 
\citep{sel_val_97, sel_14}. The S2B model studied here is the standard 
S2B model from \citet{du_etal_15}. It starts from an isolated
exponential disk that is located at the center of a rigid logarithmic
halo. The initial disk has $4\times10^6$ equal-mass particles,
softened with 0.01 length unit (the unit of length is the initial disk
scale-length). In the outer regions of the disk, the dynamical
temperature parameter (Toomre-$Q$) is roughly constant at 2.0, while
it is gradually reduced to 0.5 at the center, i.e., the central value of 
Toomre-$Q$ $b_Q=0.5$. The dynamically cool inner disk generates a bar 
instability separate from the one in the outer part, resulting in a double-barred
structure. As shown in Fig. 2 in \citet{du_etal_15}, the morphology 
of the newly formed inner bar is quite rectangular, or even peanut-like. 
After the S2B structure forms, the amplitude and morphology of the inner bar 
continue evolving, due to the interaction between the two bars, until they reach
a roughly steady state. During the steady state phase, the amplitude of 
the inner bar is relatively steady, and its morphology becomes oval-like.
The S2B structure rotates steadily in a now hotter disk, similar to a lenticular
(S0) galaxy. We analyze the kinematic properties of the S2B model in the 
steady state phase.

The scaling to physical units is obtained by setting the mass unit
$M_0=8.0\times10^{10}M_\odot$ and the length unit $R_d=2.5$ kpc, which
gives a time unit of $T_0\simeq6.6$ Myr and a velocity unit of $371$
km/s. For the S2B model, the total mass of the disk is $M_d =
1.5M_0=1.2\times10^{11}M_\odot$, extending to about $\sim 15$ kpc. The
rigid potential of the halo provides a flat rotation curve at $V_c \sim
0.6$, corresponding to 222 km/s. All analyses are made at 
$T>290\simeq1.9$ Gyr when the two bars have reached a steady 
state, during which the kinematics do not evolve much.
The rotation periods of the two bars
are stable at $P_\mathrm{inner} \sim 12.8$ $(\sim 84.5$ Myr) and
$P_\mathrm{outer} \sim 35.1$ ($\sim231.7$ Myr). Thus the inner bar
rotates roughly three times faster than its outer counterpart,
$P_\mathrm{inner}/P_\mathrm{outer} \sim 0.36$. Measured by tracing
half-way down the peak of bar amplitudes, the semi-major axes of the
outer bar and the inner bar are $a_\mathrm{outer} \sim 3.0$
($\sim 7.5$ kpc) and $a_\mathrm{inner} \sim 0.3$ ($\sim 0.75$ kpc),
respectively. For comparison purposes, we also present a single-barred
(SB) model. Using $b_Q=0.8$, the initial nuclear regions of the 
SB model are not as cool as in the S2B model, so the inner disk only 
triggers one bar
instability leading to a single bar. The bar in the SB model has a similar
semi-major axis ($a_B \sim 7.5$ kpc) and pattern speed ($P_B\sim223.1$
Myr) as the outer bar in the S2B model.

\subsection{Extracting kinematics}
\label{subsection:extkinem}
A unique advantage of simulations is that we can project the simulated 
galaxy to any desired orientation. To
extract reliable kinematics, including the high-order Gauss-Hermite
moments $h_3$ and $h_4$ \citep{van_fra_93, ger_93}, the signal-to-noise ratio 
(S/N) $\ge50$ is usually considered necessary. For simulations, given that
the number of particles follows Poissonian statistics, the required S/N can 
be transformed into a requirement on the number of particles ($N_p$) in each 
bin (S/N=$\sqrt{N_p}$). Here we apply the widely used Voronoi-binning
method \citep{cap_cop_03} to bin particles in such a way that each bin 
typically contains at least 2500 particles. Then we bin the particles 
in velocity space, with 50 velocity bins, and fit the resulting synthetic 
line-of-sight velocity distribution (LOSVD) with a Gauss-Hermite parametrization ($\bar{v}, 
\sigma, h_3$, and $h_4$). We have checked that S/N larger than 50 gives 
consistent measurements of $h_3$ and $h_4$.

\section{A general comparison of kinematic properties with the double-barred galaxies in the \atlas and \sauron \ surveys}
\label{section:Atlas3D}

\begin{figure*}[t]
       \begin{center}
       \subfigure{\includegraphics[width=0.9\textwidth]{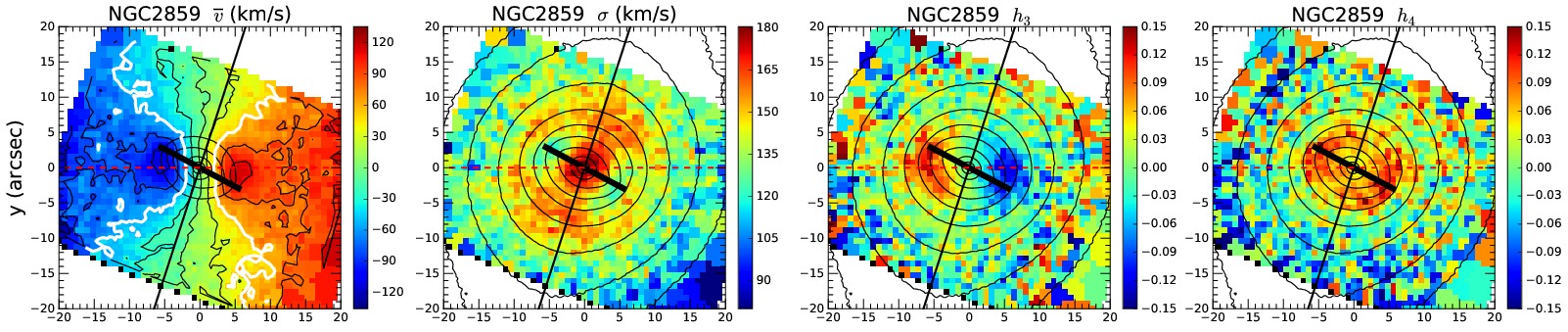}}
       \subfigure{\includegraphics[width=0.9\textwidth]{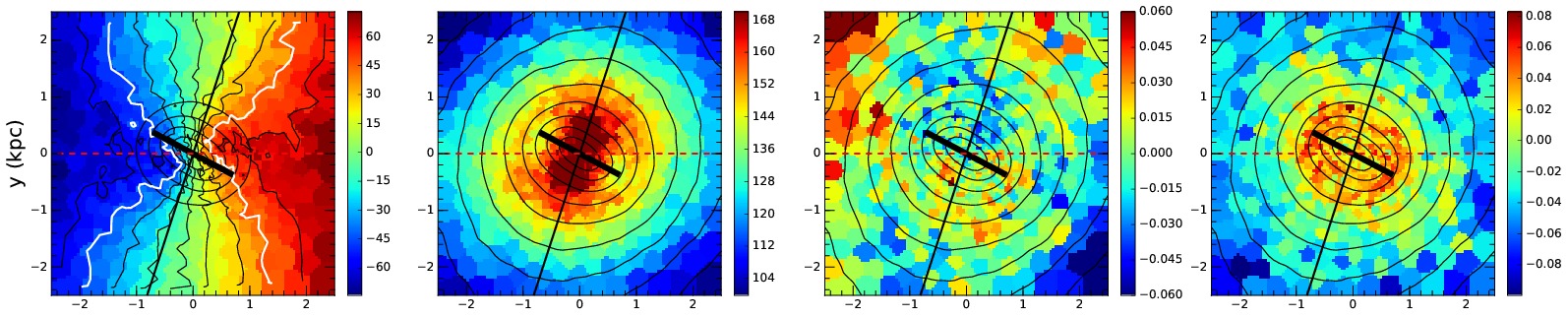}}
       \subfigure{\includegraphics[width=0.9\textwidth]{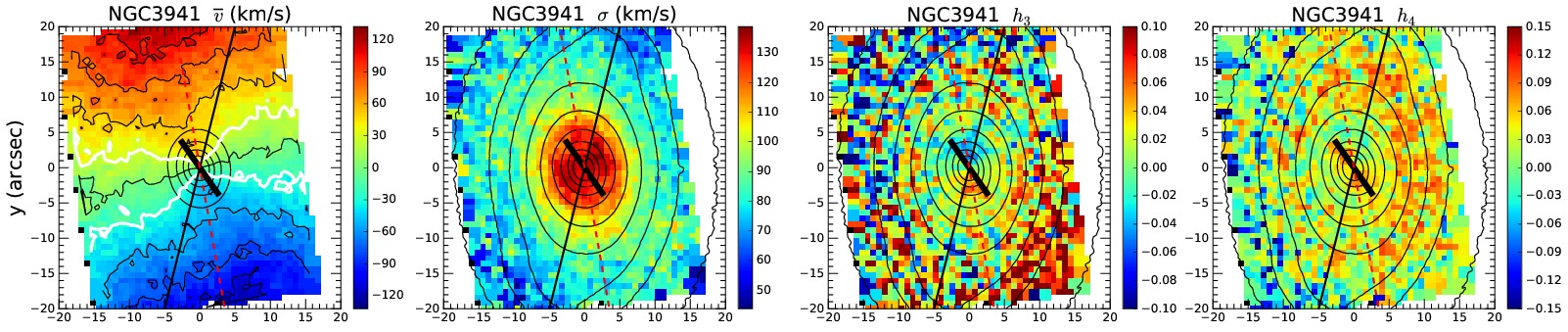}}
       \subfigure{\includegraphics[width=0.9\textwidth]{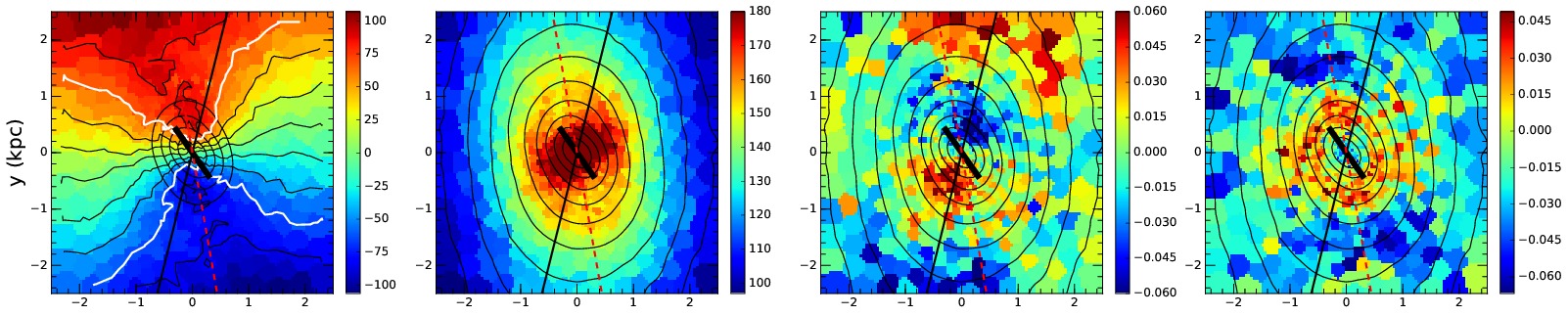}}
       \subfigure{\includegraphics[width=0.9\textwidth]{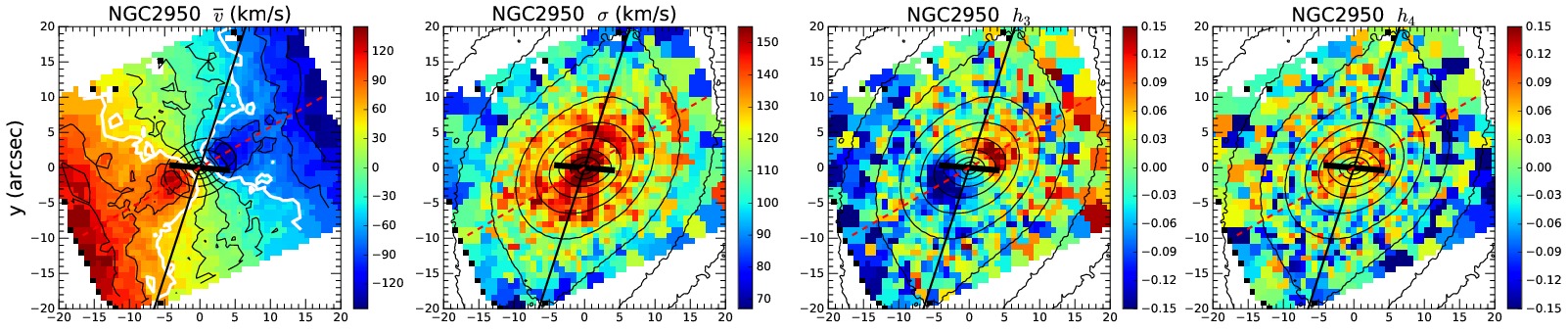}}
       \subfigure{\includegraphics[width=0.9\textwidth]{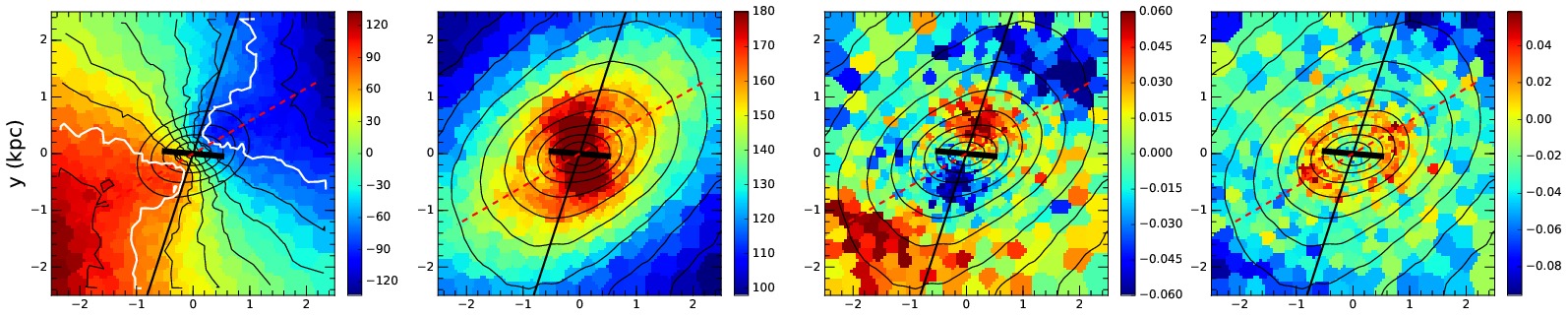}}
       \end{center}
       \caption{The kinematic maps of three S2B galaxies (NGC 2859, 3941,
       and 2950) in the \atlas \ survey \citep{cap_etal_11} followed by the best matching
     S2B model with similar orientations for the two bars and an
     identical inclination angle. 
      In all panels, the model is cropped to the same 
      area $x, y\in[-2.5, 2.5]$ kpc, in which case the size of the inner bars are similar
      in the panels presenting the model and the observations.
      The Gauss-Hermite moments of the LOSVDs are
     shown from left to right: $\bar{v}, \sigma, h_3$, and
     $h_4$. Logarithmic isodensity contours are overlaid in black. For the
     $\bar{v}$ maps, the isovelocity contours and the central
     isodensity contours are overlaid to show the twists of
     isovelocity contours caused by inner bars. A few significantly
     distorted isovelocity contours are
     highlighted with white curves. Red dashed lines indicate
     the orientations of the LON. The short black and 
     long black lines roughly show the lengths and orientations 
     of the inner and outer bars, respectively. 
      }
       \label{fig:Atlas3D}
\end{figure*}

\begin{figure*}[t]
       \begin{center}
       \subfigure{\includegraphics[width=0.9\textwidth]{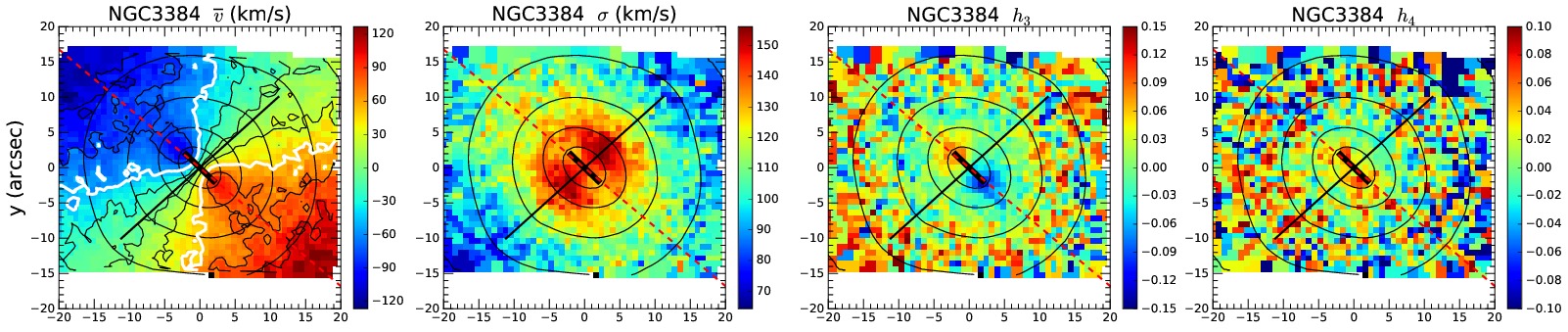}}
       \subfigure{\includegraphics[width=0.9\textwidth]{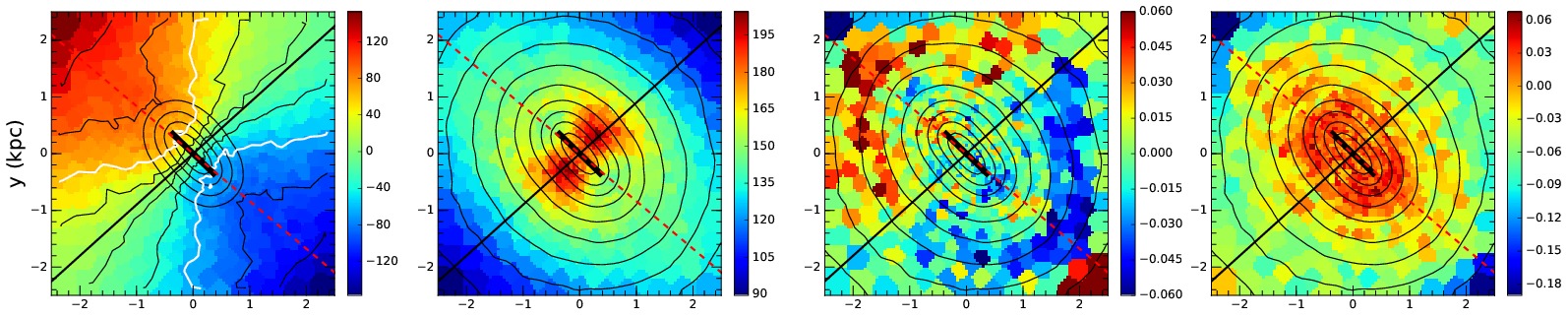}}
       \end{center}
       \caption{The kinematic maps of NGC 3384, a potential S2B candidate, obtained from 
       the \sauron \ survey \citep{ems_etal_04} followed by the best matching S2B model.} 
       \label{fig:NGC3384}
\end{figure*}
\begin{figure*}[t]
       \begin{center}
       \subfigure{\includegraphics[width=0.9\textwidth]{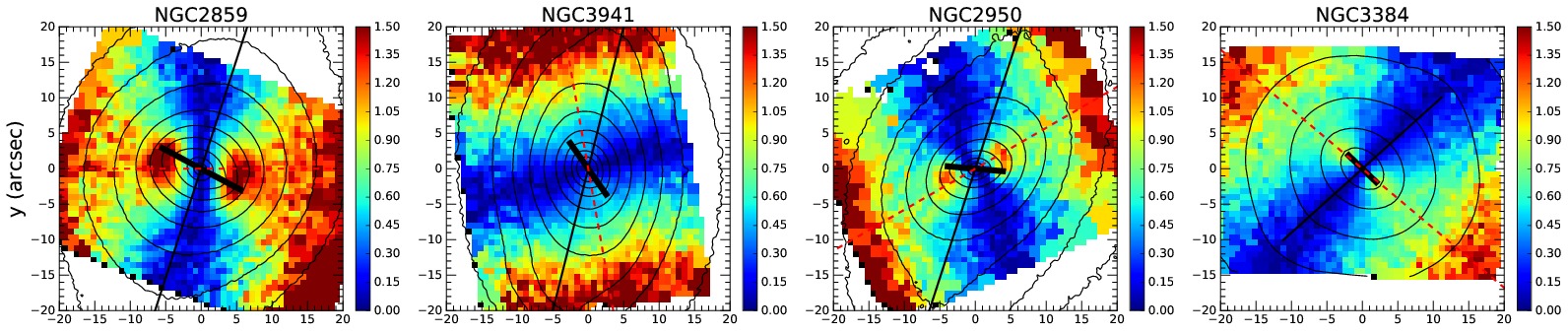}}
       \subfigure{\includegraphics[width=0.9\textwidth]{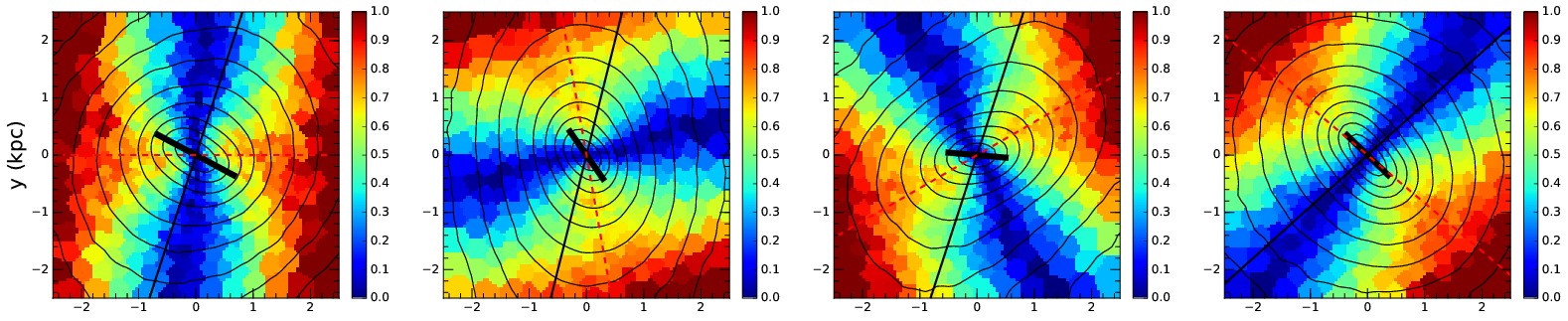}}
       \end{center}
       \caption{Maps of $\vert v_\mathrm{dp} \vert/\sigma$. From left to right: NGC 2859, 3941, 2950, and 3384, followed by the best matching S2B model.} 
       \label{fig:vsigma}
\end{figure*}

The S2B model is quite similar to massive early-type galaxies so it
can be compared with S2Bs observed in the \atlas and \sauron \ surveys. According to
the S2B catalog of \citet{erw_04} and \citet{erw_spa_03}, seven
early-type S2Bs were observed in the \atlas survey, namely \object{NGC 2859}, 
\object{NGC 2950}, \object{NGC 3941}, \object{NGC 2962}, \object{NGC 4340},
\object{NGC 3945}, and \object{NGC 7280}. The kinematics of three S2Bs
are shown in \reffig{fig:Atlas3D}. From top to bottom, they are
\object{NGC 2859}, \object{NGC 3941}, and \object{NGC 2950}, followed
by the best matching model time step. Their inner bars are well 
confirmed by ellipse fitting and unsharp mask on their infrared images. 
Based on the observed ellipticity of the outer disks, their inclination 
angles are $i\sim33^\circ, 58^\circ$, and $62^\circ$, respectively, which are 
obtained from \texttt{HyperLeda}. \object{NGC 3384} is not a well-confirmed 
S2B candidate. \citet{fis_dro_10} identified the inner bar on the infrared image 
by a photometric decomposition of the bulge region. In previous studies 
\citep{erw_spa_03, erw_04, sar_etal_06}, \object{NGC 3384} was classified as a single-barred 
galaxy, with the inner component identified as a nuclear disk by unsharp 
mask. In this paper, we consider \object{NGC 3384} as a potential S2B 
candidate, and present some kinematic evidence for the existence of the inner bar 
(\reffig{fig:NGC3384}). The inclination angle is $\sim 61^\circ$ \citep{erw_04}. 

The S2B model is inclined to the same
inclination as each target S2B at the time that the two bars have the
same relative orientation during the steady state. The short black, long black, and red
dashed lines refer to the orientations of the inner bar, outer
bar, and the line-of-nodes (LON) of the disk, respectively, as given in 
\citet{erw_spa_03} and \citet{erw_04}. The central kinematics ($R\leq20$ arcsec) 
are well resolved in all these galaxies. The logarithmically-spaced isophotes of
the S2B galaxies in \reffig{fig:Atlas3D} are plotted using the $R$-band observations of
\citet{erw_spa_03}. As shown in \reffig{fig:Atlas3D} and \ref{fig:NGC3384}, 
the S2B model qualitatively matches well with the kinematics of the observations, 
especially for $\sigma$ and $h_4$. It is worth emphasizing that only one S2B
model, at different epochs, is used to model the observations of
different galaxies. We do not expect to match these galaxies in
every aspect with such a simple model. Rather, we are interested in
the qualitative similarities between the model and observed
galaxies.

As shown in \reffig{fig:Atlas3D} and \ref{fig:NGC3384}, the most impressive kinematic 
feature is $\sigma$ enhancements, which we term ``$\sigma$-humps'', appearing along the 
minor axis of the inner bar. For \object{NGC 2859} and \object{NGC 2950}, 
$\sigma$-humps are accompanied by moderate $\sigma$ ring/spiral-like features. 
\object{NGC 3384} also exhibits $\sigma$-humps, consistent with the existence 
of an inner bar that is parallel to the LON. Relatively lower $\sigma$ at the ends 
of the inner bar can partially explain the $\sigma$-hollows found by \citet{de_etal_08}. 
We have confirmed that the initial disk does not exhibit such $\sigma$-humps/hollows, 
and the $\sigma$-humps develop with the S2B formation.

The qualitative agreement of high-order Gauss-Hermite moments is particularly 
impressive for \object{NGC 2859} and \object{NGC 2950}, especially the positive 
$h_4$ rings appearing over the projected regions of the inner bar.
These galaxies are the ones that have a velocity dispersion which is better 
resolved by the \atlas data. As noted in \citet{cap_etal_11}, for typical 
velocity dispersions $\sigma\la 120$ km/s, the LOSVD is not well resolved and 
the Gauss-Hermite moments are gradually penalized by \texttt{pPXF} \citep{cap_ems_04}
to suppress the noise in the extracted kinematics. This likely explains
the less clear structure in the $h_3$ and $h_4$ maps of \object{NGC 3941}.
Furthermore, \object{NGC 2859} and \object{NGC 2950} present clear signatures of 
fast-rotating nuclear disks, i.e., significant local maxima and minima in the $\bar{v}$ 
fields and $h_3-\bar{v}$ anti-correlations (see also \refsec{subsec:h3h4}) along 
the LON close to the center. This result is consistent with the analysis of 
\object{NGC 2859} presented in \citet{erw_etal_15} and \citet{de_etal_13}. 
To better show the relative importance of rotation and velocity dispersion, we plot 
the deprojected $|v|/\sigma$ in \reffig{fig:vsigma}. A rotation-dominated nuclear disk 
generates $|v_\mathrm{dp}|/\sigma>1$ along the kinematic major axis, where 
$v_\mathrm{dp}= \bar{v}/\mathrm{sin} i$. It is clear that both \object{NGC 2859} and 
\object{NGC 2950} also have a rotation-dominated nuclear disk within $10$ arcsec, in 
addition to the inner bar. As shown in the bottom panels of \reffig{fig:vsigma}, without 
any dynamically cold nuclear disk, the central region of the S2B model 
is significantly dominated by velocity 
dispersion, $|v_\mathrm{dp}|/\sigma \sim 0.7$. Thus in \object{NGC 2859} and \object{NGC 2950} 
the central features of $\bar{v}$ and $h_3$ are more likely to be dominated by the nuclear disk. 
It is reasonable that the S2B model does not match the $\bar{v}$ and $h_3$ fields perfectly.
\object{NGC 3384} shows moderate rotation in the central region, 
$|v_\mathrm{dp}|/\sigma \sim 0.8$, thus it is unclear whether a nuclear disk exists or not. 

In conclusion, the S2B model is able to qualitatively match many of the kinematics of 
observed S2Bs, making it very useful for studying the kinematics of S2Bs. In the following
section, we use this model to analyse these kinematic properties in detail.

\section{Kinematic analyses}
\label{section:kinem}

\subsection{$\bar{v}$ twists}

\begin{figure*}[htp]
        \centering
        \includegraphics[width=0.95\textwidth]{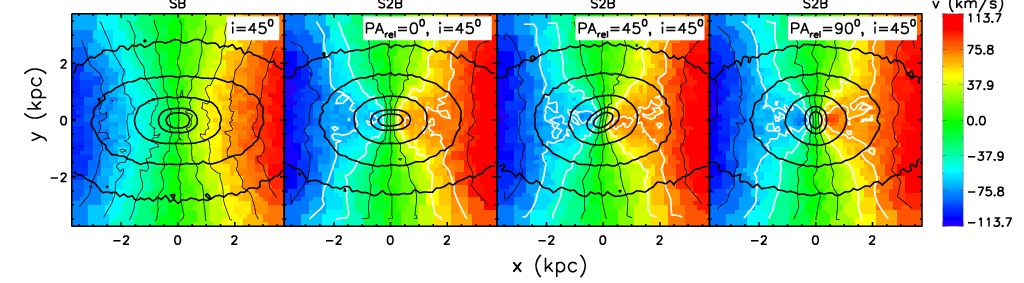}
        \caption{The $\bar{v}$ maps of the SB model (the
          leftmost panel) and the S2B model, showing the twists 
          caused by a rapidly rotating inner bar. Highlighted with
          white contours, the $\bar{v}$ contours show that the most
          significant twists are present at an intermediate
          region between the two bars. The outer bar is fixed along the
          LON ($x$-axis). For the S2B model, the relative angle
          PA$_\mathrm{rel}$ between the inner bar and the outer bar is
          $0^\circ, 45^\circ$, and $90^\circ$, from left to right. The
          disk is inclined at $i=45^\circ$. The logarithmic isodensity
          contours (levels 0.9, 0.8, 0.6, 0.4, 0.2 of 
          ln$(\Sigma_\mathrm{max} / \Sigma_\mathrm{min})$) are overlaid in black. The
          contours at levels 0.2 and 0.8 roughly coincide with the
          photometric edge of the outer bar and the inner bar,
          respectively.}
        \label{meanv}
\end{figure*}

In \reffig{meanv}, we show $\bar{v}$
fields for the S2B and SB models. Aligned with the LON (here the
$x$-axis), the large-scale bar in the SB model shows smooth and nearly
parallel isovelocity contours in the region of the bar (the
leftmost panel, inclination $i=45^\circ$), as expected. The other
three panels show $\bar{v}$ fields of the S2B model when the relative
position angle of the two bars (PA$_{\rm rel}=\vert$ PA$_{\rm inner} -
$PA$_{\rm outer}\vert$) is $\sim 0^\circ, 45^\circ$, and $90^\circ$,
respectively. For the S2B model, $\bar{v}$ fields are similar to 
those of the SB model in the large-scale bar regions, as the large-scale 
bars in the S2B and SB models rotate at nearly the same pattern speed. 
At the very central regions where the inner bar 
dominates, the isovelocity contours even break up, forming local minima and maxima, 
especially the inner bar is perpendicular to the LON (the 
rightmost panel, PA$_{\rm rel}=90^\circ, i=45^\circ$). 
In the case of an inclined axisymmetric nuclear disk, the kinematic 
axis must align with the LON. Since stars in bar regions have significant 
radial streaming motions, the kinematic axis is expected to be misaligned 
from the LON. As shown in \reffig{meanv} when PA$_\mathrm{rel}=45^0$, within 
the projected regions of the inner bar, the isovelocity contours are slightly 
distorted toward the opposite direction of the major axis of the inner 
bar, in agreement with \citet{she_deb_09}. Thus it might be because the 
twists are too weak that the kinematic axis of the observed galaxies 
(\reffig{fig:Atlas3D} and \ref{fig:NGC3384}) does not exhibit a clear misalignment 
with the LON in the central regions. Furthermore, as presented in \refsec{section:Atlas3D}, 
\object{NGC 2859} and \object{NGC 2950} have a rotation-dominated nuclear disk which 
also significantly weakens the misalignment. Therefore, it is not surprising that 
previous observations \citep{moi_etal_04} did not find clear signs of the existence 
of decoupled inner bars in the form of central velocity twists.
Moreover, we also did not find significant kinematic misalignments at the central 
regions of the single-barred (SB) model at most inclinations and bar orientations, 
which is consistent with the IFU observations of early-type barred galaxies 
\citep{cap_etal_07, kra_etal_11}. The IFU observations show that the kinematic 
misalignment is always quite small ($\sim5^\circ$ level). We propose that such a 
result is because the bars in early-type galaxies have large random motions.

Although the velocity twists are insignificant in the central regions, 
we notice that the observed S2Bs present consistent velocity twists at intermediate 
radii between the two bars. As shown in \reffig{meanv}, for the S2B model, 
some isovelocity contours (highlighted with white curves) are significantly distorted 
toward the central regions at intermediate radii ($R\sim1.5$ kpc) of 
the transition zone between the two bars. Independent of the relative orientation 
of the two bars and the inclination, such twists occur far from the photometric 
ends (the isodensity contour at level 0.8) of the inner bar. 
The transition zone where $\bar{v}$ twists occur is the 
region where the two bars are mixed and interacting.
For \object{NGC 2859}, and \object{NGC 2950}, significant twists also appear at
positions quite far from the end of the inner bar (highlighted
with white curves), which is consistent with the S2B model. It is 
also worth noticing that the twists can be significantly asymmetric with 
respect to the LON. For example, the asymmetric twists in \object{NGC 2950} 
can be clearly seen along the white isovelocity contours. 
On the upper side of the LON, the top-left arm is slightly distorted, 
while, on the lower side, the bottom-left arm has a nearly $90^\circ$ 
twist. The S2B model gives twists very consistent with those in \object{NGC 2950}. 
\object{NGC 3941}, which hosts only a weak inner bar, 
does not exhibit twists as significant as the model in their $\bar{v}$ fields, 
and the weak twists occur at positions close to the end of the inner bar. 
Without nuclear disk in the model, such asymmetric twists may 
be caused by the non-axisymmetric motions in the transition zone of the S2B structure.

\subsection{$\sigma$-humps}
\label{subsec:sigma}
\begin{figure*}[htp]
    \centering
        \subfigure{\includegraphics[width=0.85\textwidth]{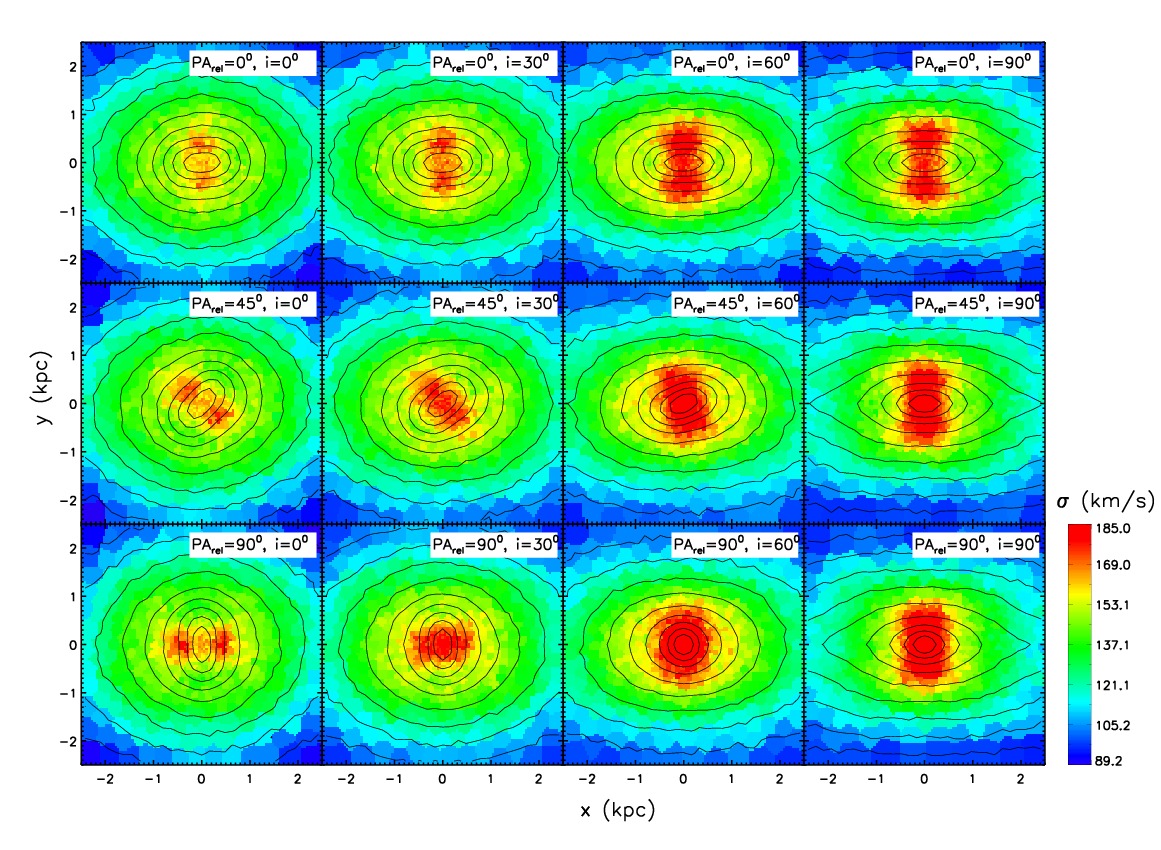}}
        \caption{The line-of-sight velocity dispersion of the S2B
          model. From top to bottom, the relative orientations between
          bars are PA$_{\rm rel}=0^\circ, 45^\circ$, and $90^\circ$,
          respectively. The inclination varies from $0^\circ$ to
          $90^\circ$, left to right. The outer bar and LON are fixed
          on the $x$-axis. Logarithmically spaced isodensity contours
          are overlaid in black.}
        \label{fig:S2Bsigma}
        \subfigure{\includegraphics[width=0.85\textwidth]{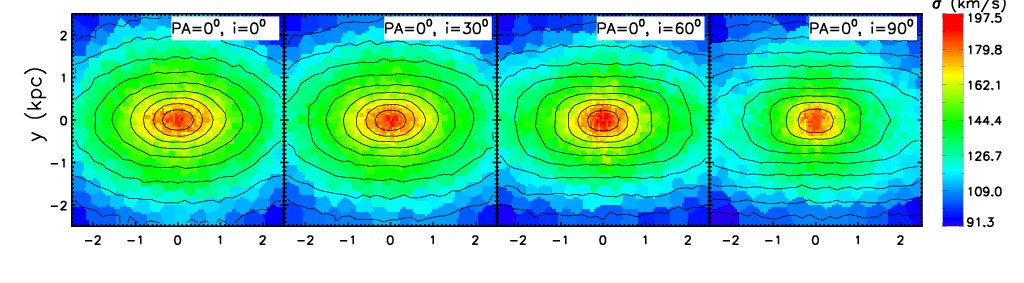}}
        \subfigure{\includegraphics[width=0.85\textwidth]{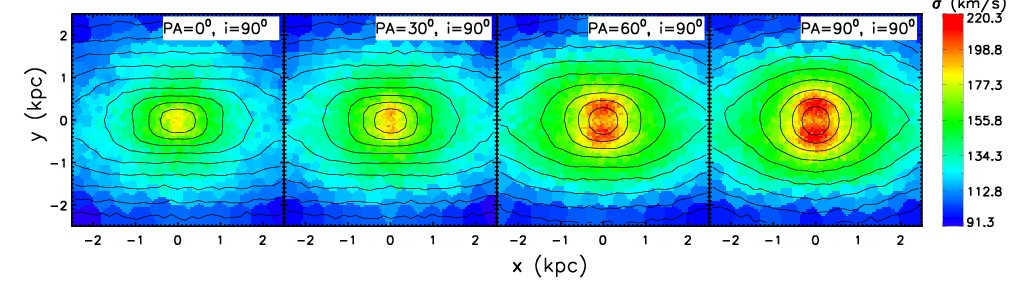}}
        \caption{The line-of-sight velocity dispersion of the SB
          model. For the top panels, the bar is aligned with the LON ($x$-axis); the 
          inclination angle $i$ varies from $0^\circ$ to $90^\circ$. The bottom panels show 
          the edge-on view of the SB model when the position angle of the bar varies from 
          $0^\circ$ to $90^\circ$ with respect to the LON.}
        \label{fig:PBsigma}
\end{figure*}

\begin{figure*}[htp]
       \centering
       \subfigure{\includegraphics[width=1.0\textwidth]{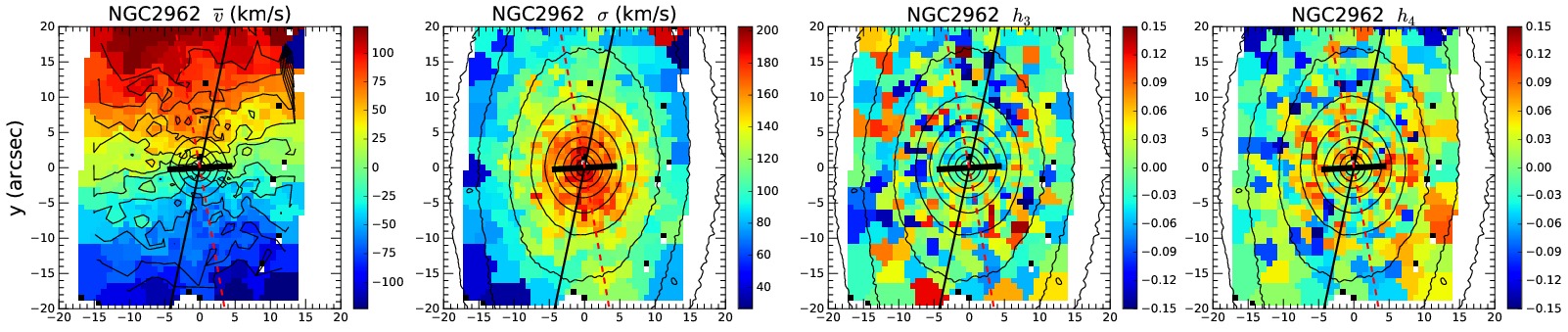}}
       \subfigure{\includegraphics[width=1.0\textwidth]{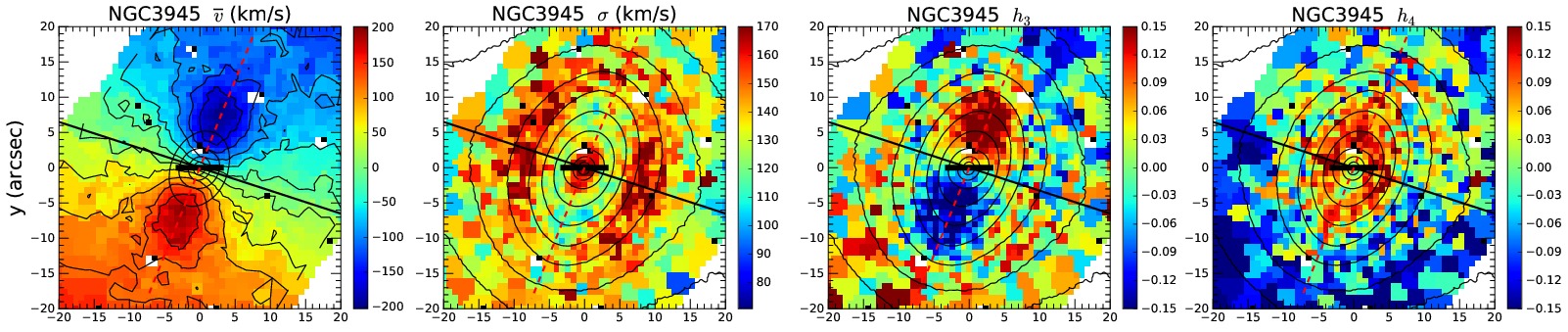}}
       \subfigure{\includegraphics[width=1.0\textwidth]{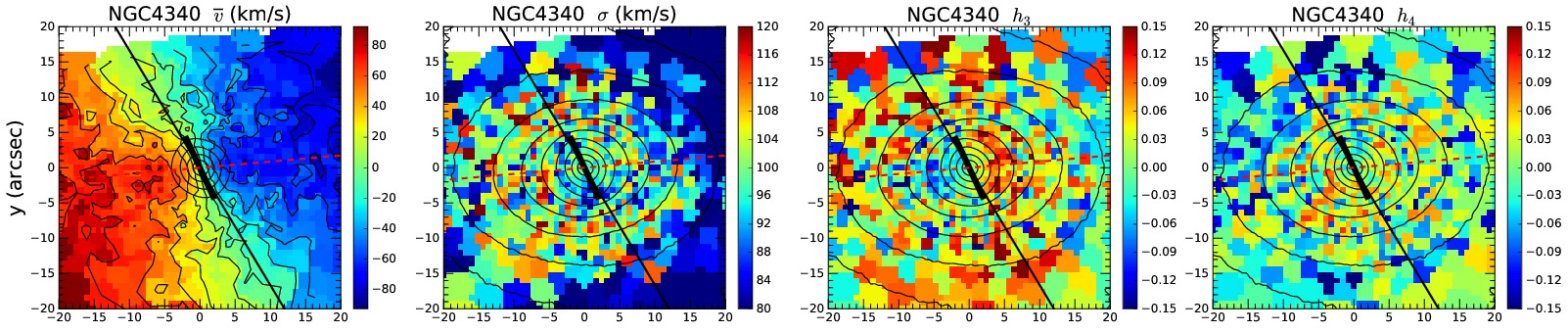}}
       \subfigure{\includegraphics[width=1.0\textwidth]{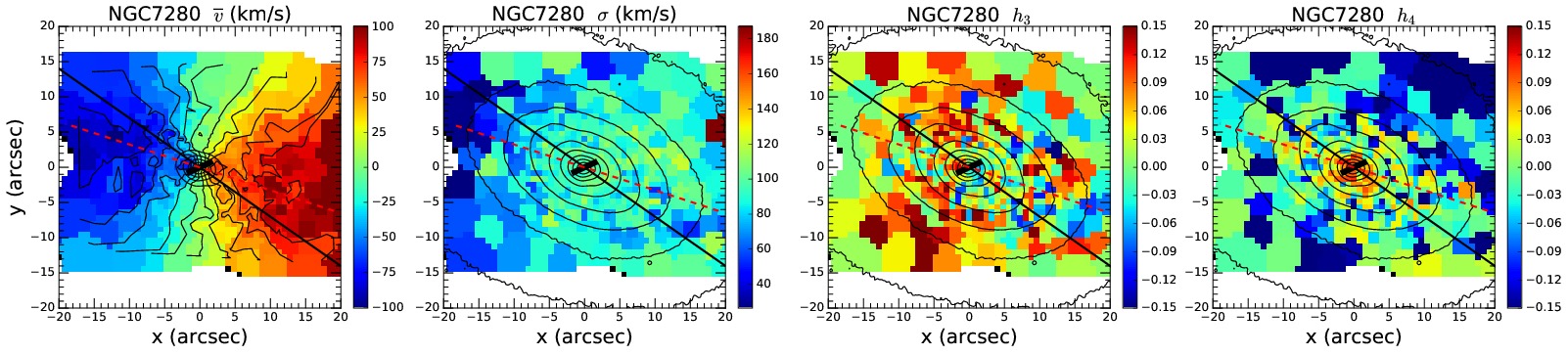}}
       \caption{The kinematic maps for the remaining four S2B galaxies
         (NGC 2962, 3945, 4340, and 7280) in the \atlas \ survey which do not show significant 
         $\sigma$-humps.}
       \label{fig:Atlas3Dno}
\end{figure*}

\subsubsection{Basic properties}

The $\sigma$ maps of the S2B model are shown in
\reffig{fig:S2Bsigma}. Because varying the orientation of the outer
bar does not lead to a significant difference, we fix the outer bar
on the LON (the $x$-axis). The relative position angles of the two
bars PA$_{\rm rel}$ are $\sim 0^\circ$, $45^\circ$, and $90^\circ$,
respectively, from top to bottom. From left to right, the disk is
inclined from $i=0^\circ$ to $i=90^\circ$. In most cases, significant
$\sigma$-humps appear along the minor axis of the inner bar. In
contrast, in the SB model the elliptical $\sigma$ peak is aligned with
the bar (the top row in \reffig{fig:PBsigma}), which is
consistent with previous single-barred models \citep{mil_smi_79,
vau_dej_97}, except for edge-on views.

As shown in \reffig{fig:S2Bsigma}, $\sigma$-humps are closely associated with 
the inner bar. The basic properties of $\sigma$-humps are elaborated below. 
1) $\sigma$-humps have a similar size as the inner bar. 
2) The $\sigma$-humps are always present on the minor axis of the
inner bar. In long-slit measurements, such $\sigma$-humps should
appear as double-peaked or flat-topped distributions along the minor
axis, while along the major axis $\sigma$ is relatively lower, forming
$\sigma$-hollows. The difference can be a few tens of km/s in this S2B
model. 3) Projected properties of $\sigma$-humps are significantly
affected by the orientation of the inner bar and the
inclination of the disk. In most cases, $\sigma$-humps are clearly
visible. However, when the inner bar is nearly perpendicular to
the LON, $\sigma$-humps are barely visible at intermediate
inclinations ($i\sim40^\circ-70^\circ$). Then the difference in $\sigma$
between the minor and major axes becomes small, giving a nearly
axisymmetric $\sigma$ distribution. 4) Around $\sigma$-humps/hollows,
rings of enhanced $\sigma$ (by $\sim 10$ km/s) are 
present at most inclinations and orientations, except for edge-on views. 
5) The amplitude of $\sigma$-humps oscillates in
a similar way as the inner bar, i.e., $\sigma$-humps are weaker
when the two bars are parallel (PA$_{\rm rel}=0^\circ$) and stronger
when the bars are perpendicular (PA$_{\rm rel}=90^\circ$). 
The physical origin of $\sigma$-humps is still unclear. \citet{de_etal_08} proposed
that $\sigma$-hollows might originate from a dynamically cold inner bar which is embedded in
a relatively hotter (classical) bulge. Then the $\sigma$-humps/hollows are generated
by the contrast between the hotter bulge and the cold bar. The S2B model
presented here does not contain any classical bulge, but a B/P bulge, forming
from the internal buckling instability of the outer bar, which is also hotter
than the inner bar. Further studies are needed to clarify whether $\sigma$-humps can be
explained in terms of the contribution of bulges, as suggested by \citet{de_etal_08}.
      
Regardless of the relative orientation of the two bars, significant
vertically extended $\sigma$ features (termed ``vertical $\sigma$-humps'') 
are present in edge-on views of both the S2B model (the rightmost 
panels of \reffig{fig:S2Bsigma}) and the SB model (the bottom panels of 
\reffig{fig:PBsigma}). For the S2B model, the vertical $\sigma$-humps
in edge-on views are more extended and have larger values than $\sigma$-humps 
in face-on views. In the side-on view of the inner bar (PA$_{\rm
rel}=0^\circ$, $i=90^\circ$), $\sigma$ becomes lower close
to the mid-plane than in the end-on view (PA$_{\rm rel}=90^\circ,
i=90^\circ$). For the SB model, with increasing position angle of the 
bar from $0^\circ$ (side-on) to $90^\circ$ (end-on), the vertical $\sigma$-humps 
become more pronounced in the projected bar regions. \citet{ian_ath_15} 
found similar features in edge-on views of their B/P bulge models hosting a 
single bar. \citet{qin_etal_15} also found a similar vertically
extended $\sigma$ feature in their Milky Way bar model \citep{she_etal_10}. 
\citet{fal_etal_06} found that three of their five nearly edge-on
galaxies (\object{NGC 3623}, \object{NGC 4235}, and \object{NGC 5689}) 
show such vertical $\sigma$-humps. In the \atlas survey \citep{cap_etal_11}, we also find 
some edge-on galaxies exhibiting significant vertical $\sigma$-humps 
(\object{NGC 2549}, \object{3301}, \object{3610}, \object{4026}, \object{4111}, 
\object{4251}, \object{4342}, \object{4417}, \object{5308}, 
\object{5322}, and \object{5422}). It is clear that both observations and simulations 
suggest that the vertical $\sigma$-humps are very common in edge-on 
galaxies. As suggested by \citet{ian_ath_15}, the peak value and extension of 
vertical $\sigma$ enhancements may be monotonic with the strength of the B/P bulge. 
However, as shown in \reffig{fig:S2Bsigma} and \ref{fig:PBsigma}, both our S2B 
and SB models indicate that the existence of a bar can also significantly affect 
the properties of such vertically extended $\sigma$ features. But, because of the projection, 
it is hard to study the relation between bars and vertical $\sigma$-humps in 
the real edge-on galaxies. The physical origin and the relation with bars of such 
vertical $\sigma$ humps are still not clear; the vertical $\sigma$ humps seem not to 
have the same origin as the $\sigma$-humps appearing in face-on views.

As shown in \reffig{fig:Atlas3D} and \ref{fig:NGC3384}, three S2Bs and one S2B candidate 
in the \atlas and \sauron \ surveys show $\sigma$-humps, and are well-matched by the
model. \object{NGC 2859} and \object{NGC 3941} were included in the
sample of \citet{de_etal_08}. For \object{NGC 2859}, the model also
exhibits a similar moderate $\sigma$-ring feature around the
$\sigma$-humps, as in the observation, which causes significant
$\sigma$-hollows appearing at the ends of the inner bar. The
$\sigma$-humps in \object{NGC 3941} are not as significant as in the
model, probably because the inner bar in \object{NGC 3941} is much
weaker than in the model, with rounder isodensity contours and a
smoother $\bar{v}$ field. \object{NGC 2950} and \object{NGC 3384} are 
new examples of $\sigma$-hollow/hump galaxies. 
The orientation of the $\sigma$-humps in \object{NGC 2950} is not accurately consistent with the model, which might
be caused by the differences in their kinematic details between NGC 2950 and the numerical model.
It is worth noticing that, around the $\sigma$-humps in \object{NGC 2950} and the model, there are diffuse $\sigma$ spiral-like
features which might be similar with the $\sigma$ ring in \object{NGC 2859}.
Using the Voronoi binning method with high enough S/N, such $\sigma$ ring/spiral-like features are statistically
significant kinematic features which are still poorly understood.
In conclusion, based on the simulation, we argue that $\sigma$-hollows are the
same feature as $\sigma$-humps accompanied by $\sigma$ ring/spiral-like 
features sometimes, viewed differently.

Finally, we briefly discuss the reason why the other four galaxies
show no significant $\sigma$-humps. As shown in
\reffig{fig:Atlas3Dno}. \object{NGC 2962} and \object{NGC 4340} are
intermediately inclined, and their inner bars are almost
perpendicular to the LON, $83^\circ$ and $70^\circ$, respectively. The
model shows that the $\sigma$-humps are less noticeable in such
conditions. For \object{NGC 3945}, as mentioned above, the inner regions are dominated
by a large nuclear disk which significantly affects the kinematic
properties \citep{erw_etal_03, col_etal_14}, leaving no clear
kinematic signatures of the inner bar. Finally, the inner bar
in \object{NGC 7280} seems too short to generate distinguishable
kinematics. The S2B model may not match these galaxies very
well.

\subsubsection{$\sigma$-humps in aligned double-barred galaxies}

\begin{figure}[htp]
        \centering
        \includegraphics[width=0.48\textwidth]{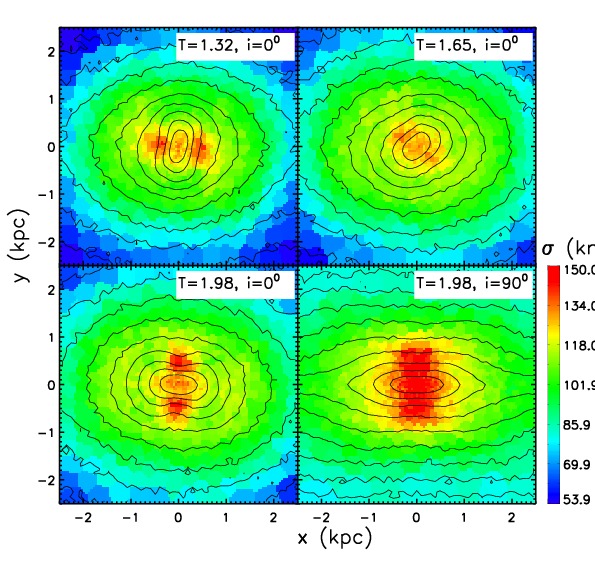}
        \caption{The co-evolution between $\sigma$-humps and the
          inner bar in an aligned S2B. The coupling process
          happens during $T=1.65-1.98$ Gyr. The peanut-shaped relic of
          the inner bar and $\sigma$-humps exist until the end of
          this simulation at $T\simeq 4.0$ Gyr.}
        \label{sigmacouple}
\end{figure}
We have explored the kinematics of the simulations reported in
\citet{du_etal_15}. We find that $\sigma$-humps are not unique
features of S2Bs. They are also present in galaxies hosting a single
small-scale bar (i.e., nuclear-barred galaxies), and also aligned
S2Bs, where the two bars have coupled into alignment leaving only a single
bar. In \citet{du_etal_15}, we showed that the coupling process
distorts the iso-density contours to a peanut shape, which may be used
to distinguish aligned S2B galaxies from normal single-barred
galaxies. We plot the evolution of $\sigma$-humps during the coupling 
process in \reffig{sigmacouple}. This aligned S2B model has the same 
initial conditions as the model in Fig. 11 in \citet{du_etal_15}, but its 
stellar mass ($M_d=1.0$) is slightly lower. Before coupling, as
shown in the top panels, the amplitude of the $\sigma$-humps gradually
decrease with the strength of the inner bar. After the inner
bar is trapped ($T\ge1.98$ Gyr, the bottom panels) by its outer
counterpart, the peanut-shaped relic of the inner bar exhibits 
significant $\sigma$-humps on its minor axis. The co-evolution with
the inner bar indicates that $\sigma$-humps may be used as a
diagnostic of aligned S2Bs. We also expect that the amplitude of
$\sigma$-humps will be affected by the bar strength in different
models.

\subsection{Higher-order moments: $h_3$ and $h_4$}
\label{subsec:h3h4}

\begin{figure*}[htp]
    \centering
    \includegraphics[width=0.95\textwidth]{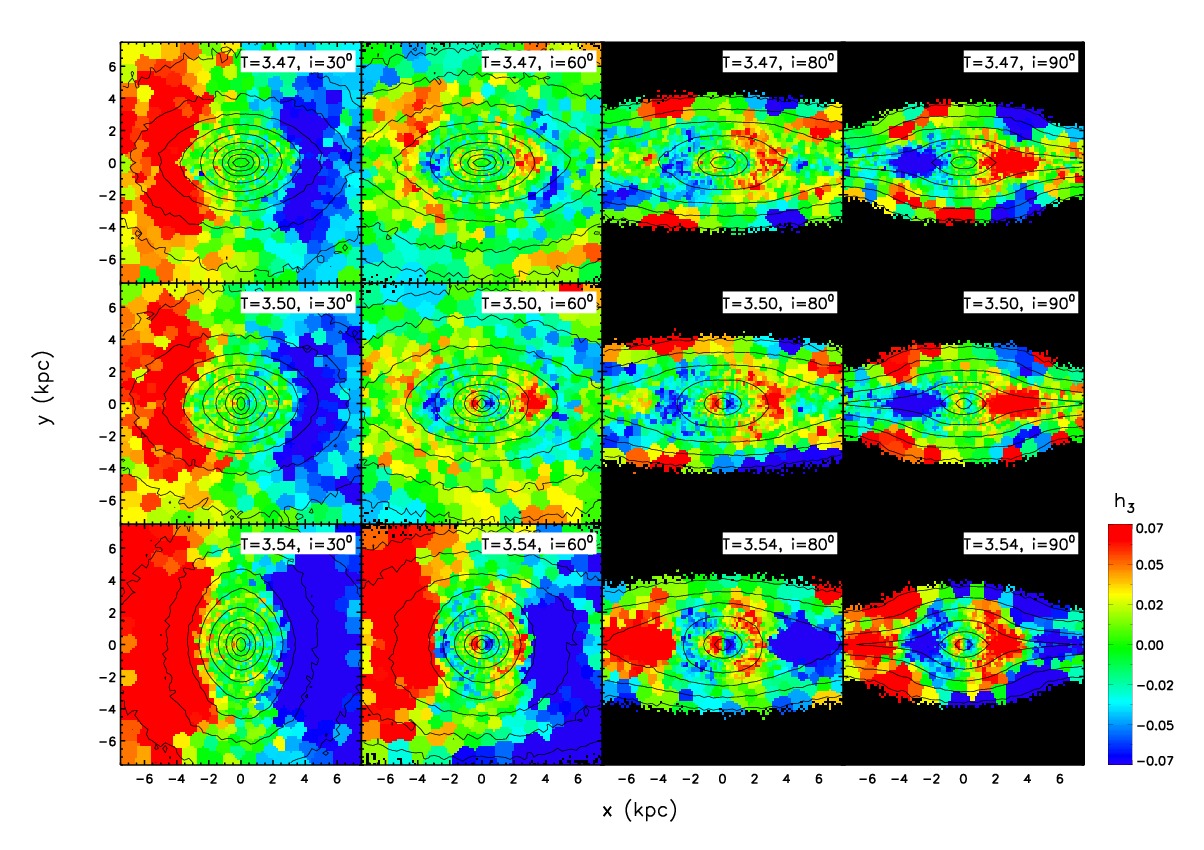}
    \caption{$h_3$ maps varying the orientation and inclination of the
      bars.  The LON is fixed on the $x$-axis. From left to right:
      $i=30^\circ, 60^\circ, 80^\circ$, and $90^\circ$. $\bar{v}$ is 
      negative at $x<0$ and positive at $x>0$. Isodensity
      contours are overlaid in black to show the orientation of two bars. At
      $T=3.50$ and $3.54$ Gyr, the inner bar is perpendicular to
      the LON, while it is parallel to the LON at $T=3.47$ Gyr. The
      outer bar is perpendicular to the LON at $T=3.54$ Gyr, but
      parallel to the LON at $T=3.47$ and $3.50$ Gyr. The color is set
      to black when the number of particles in one pixel
      ($\sim150\times150\ \mathrm{pc}^2$) is less than 20. Such
      regions cannot collect enough particles even with the Voronoi binning,  
      and thus do not provide reliable kinematics.}
    \label{fig:S2Br3h3}
\end{figure*}

\begin{figure*}[htp]
        \centering
        \includegraphics[width=0.95\textwidth]{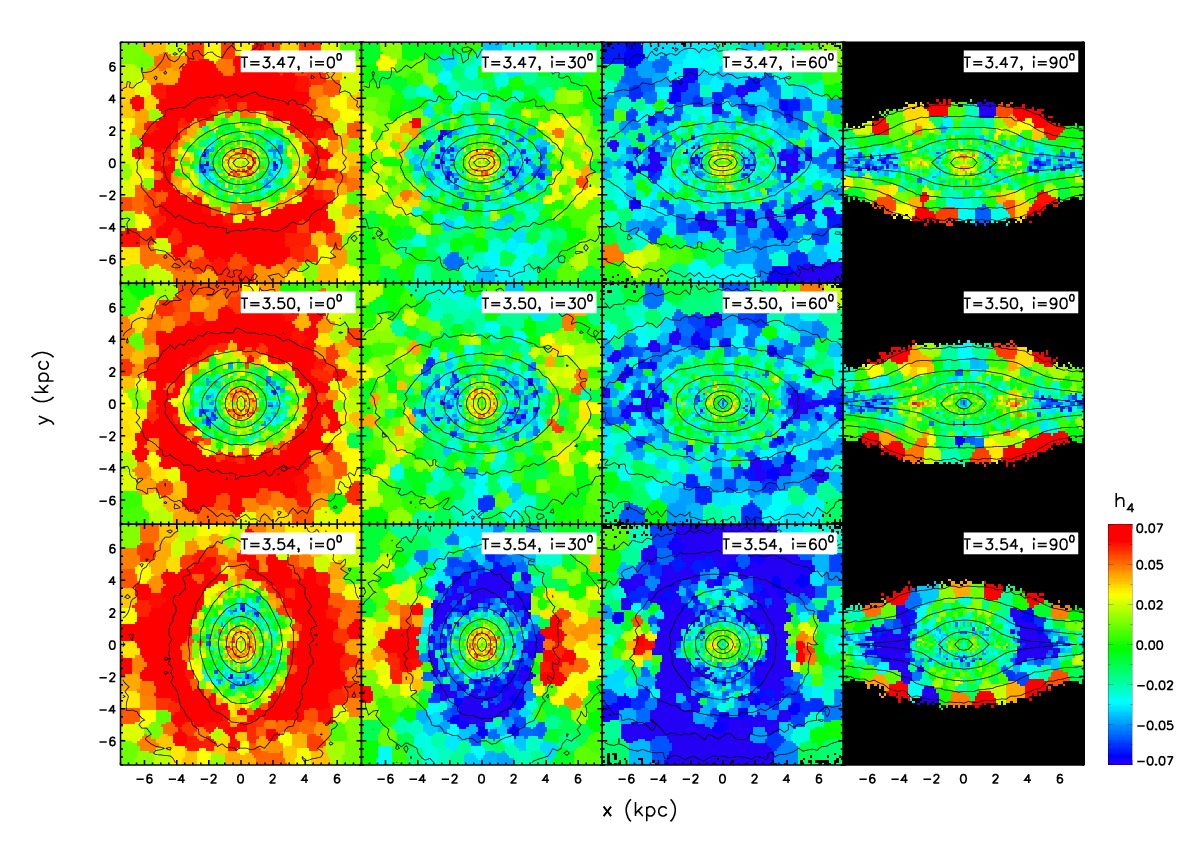}
        \caption{$h_4$ maps varying the orientations of the two bars
          and the inclination angle. The orientations of the LON and
          the two bars are the same as \reffig{fig:S2Br3h3}. From left
          to right, the disk is inclined at $i=30^\circ, 60^\circ,
          80^\circ$, and $90^\circ$, respectively. Isodensity contours
          are overlaid in black.}
        \label{fig:S2Br3h4}
\end{figure*}

The higher-order Gauss-Hermite moments $h_3$ and $h_4$ describe the
asymmetric and symmetric deviations, respectively, from a pure
Gaussian \citep{ger_93, van_fra_93, ben_etal_94}. It is well known
that the LOSVD of an axisymmetric disk generally has a lower velocity
tail due to the projected outer disk, yielding an anti-correlation of
$h_3$ and $\bar{v}$. For barred galaxies, \citet{bur_ath_05} showed
that in the bar regions, $h_3$ becomes correlated with $\bar{v}$ in
edge-on views. They presented that the $h_3-\bar{v}$ correlation is an
indication of the high-velocity tail created by the elongated orbits
supporting the bar. The fourth-order Gauss-Hermite coefficient $h_4$
is negative when a distribution is broader than Gaussian and positive
when it is more peaked. \citet{deb_etal_05} showed that a B/P-shaped
bulge generates a flat-topped LOSVD in face-on views.

In \reffig{fig:S2Br3h3}, we present $h_3$ maps for the S2B model that
cover the whole region of the outer bar ($\sim7.5$ kpc). The disk is
inclined to $i=30^\circ, 60^\circ, 80^\circ$, and $90^\circ$ with
respect to the $x$-axis. $\bar{v}$ is negative at $x<0$ and 
positive at $x>0$. It
is not surprising that widespread $h_3-\bar{v}$ anti-correlation
appears around the outer bar where LOSVDs are dominated by circular
motions of the disk. With increasing $i$, the correlation between
$h_3$ and $\bar{v}$ expands to all over the projected regions of the
outer bar. This result is consistent with the single-barred cases studied
in \citet{bur_ath_05}. Closer to the center, the inner bar
presents its own $h_3$ features. When the inner bar is
nearly perpendicular to the LON (e.g. $T=3.50, 3.54$ Gyr),
$h_3-\bar{v}$ changes to anti-correlation again at high inclinations
$40^\circ \le i \le 90^\circ$, while when the inner bar is
parallel to the LON (e.g. $T=3.47$ Gyr) there is no significant
$h_3-\bar{v}$ anti-correlation. 
We propose that the $h_3-\bar{v}$ anti-correlation, appearing in the projected inner bar regions, is caused by the significant
streaming motions in the inner bar which compose the high velocity peak of the LOSVDs, thus generating
$h_3-\bar{v}$ anti-correlation. In order to clarify the $h_3$ features clearly, we considered the non-Gaussian LOSVDs as
bimodal profiles which are assumed to be composed by two independent Gaussian components, i.e., the high $\vert \bar{v} \vert$ component and the low
$\vert \bar{v} \vert$ component. Used to quantify the asymmetric deviation, $h_3$ is equal to zero in the cases that the two components
have the exactly same distribution. In the cases that the low $\vert \bar{v} \vert$ component is stronger than the high
$\vert \bar{v} \vert$ component, the LOSVD is composed by a main peak dominated by low $\vert \bar{v} \vert$ component and a
high $\vert \bar{v} \vert$ tail, in which case $h_3$ is correlated with $\bar{v}$. In contrast, if the high $\vert \bar{v} \vert$
component is stronger, the LOSVD is composed by a high $\vert \bar{v} \vert$ peak and a low $\vert \bar{v} \vert$ tail, yielding an
anti-correlation of $h_3$ and $\bar{v}$. In the outer bar regions, the high-speed streaming motions in the outer bar generate the high
$\vert \bar{v} \vert$ tail in the LOSVDs which peak at low $\vert \bar{v} \vert$, thus generating $h_3-\bar{v}$ correlation. In the
inner bar regions, because a large fraction of stars participate in the high-speed streaming motions in the inner bar, the high
$\vert \bar{v} \vert$ component dominates the LOSVDs, thus $h_3$ can be anti-correlated with $\bar{v}$. Especially when the inner
bar is perpendicular to the LON, the elongated motions in the inner bar well overlap with the line-of-sight to some extent,
leading to most prominent $h_3-\bar{v}$ anti-correlations. Because at $i=90^\circ$ the bars and the disk are superposed, the
nuclear $h_3-\bar{v}$ anti-correlation is weakened.

As shown in \reffig{fig:S2Br3h4}, the S2B also has noticeable $h_4$
features. In nearly face-on views, the most impressive $h_4$ feature
is positive rings around the outer and the inner bars. Such positive
$h_4$ rings around the inner bar were also noted by
\citet{she_deb_09}. With increasing inclination, positive $h_4$ rings
become weaker and gradually disappear. Such features suggest that,
compared with their surroundings, bars are more tightly bound at the
mid-plane, thus the vertical velocity distribution is more peaked. In the
intermediate regions, the negative $h_4$ probably corresponds to the
B/P-shaped bulge.

We can also see complex non-Guassian features around the inner
bars in observed S2Bs (\reffig{fig:Atlas3D} and \ref{fig:NGC3384}), 
i.e., $h_3-\bar{v}$ anti-correlations and positive $h_4$ rings. $h_3-\bar{v}$
anti-correlations have also been considered as a tracer of nuclear 
disks \citep{bur_ath_05}. In all these S2Bs, anti-correlated $h_3$
are roughly aligned with the LON, while $h_3$ features have
significant misalignments in the model. As mentioned above, both
\object{NGC 2950} and \object{NGC 2859} probably host a nuclear disk
that generates $h_3$ features well aligned with the LON. Therefore,
though we cannot decompose nuclear disks and inner bars, the
inner bar does provide an alternative explanation of nuclear
$h_3-\bar{v}$ anti-correlations. The inner bars show positive
$h_4$ rings in \object{NGC 2859} and \object{NGC 2950}, which are
consistent with the model, although the absolute value is much larger
in the observations. \object{NGC 3384} also shows some positive 
$h_4$ features and $h_3-\bar{v}$ anti-correlation at the bar regions, 
which are roughly consistent with the S2B model. For \object{NGC 3941}, 
positive $h_4$ is widely distributed all over the disk. At the central 
regions where the resolution is highest, we cannot identify a
$h_4$ ring that is closely related to the inner bar, possibly 
because of the penalization on high-order Gauss-Hermite moments
with the \ppxf \ method, as mentioned in \refsec{section:Atlas3D}.

\subsection{Intrinsic kinematics}
\begin{figure*}[htp]
    \centering
    \subfigure{\includegraphics[width=1.0\textwidth]{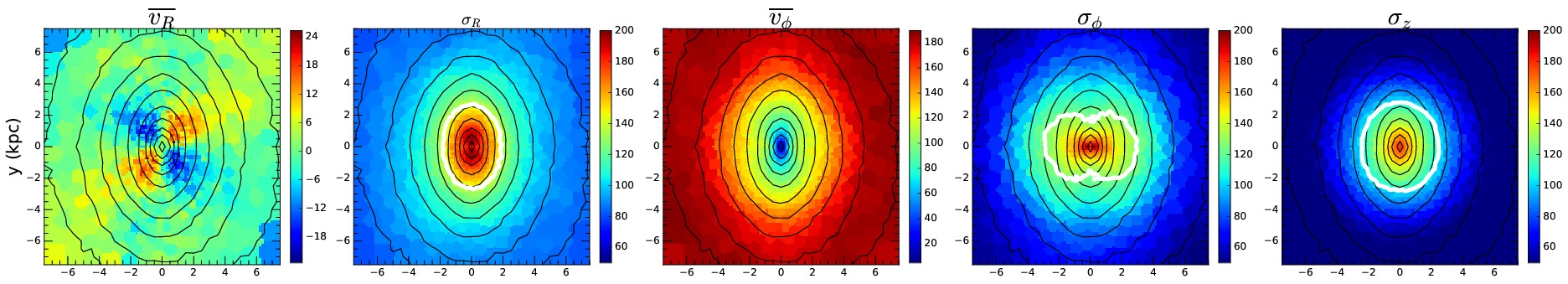}}
    \subfigure{\includegraphics[width=1.0\textwidth]{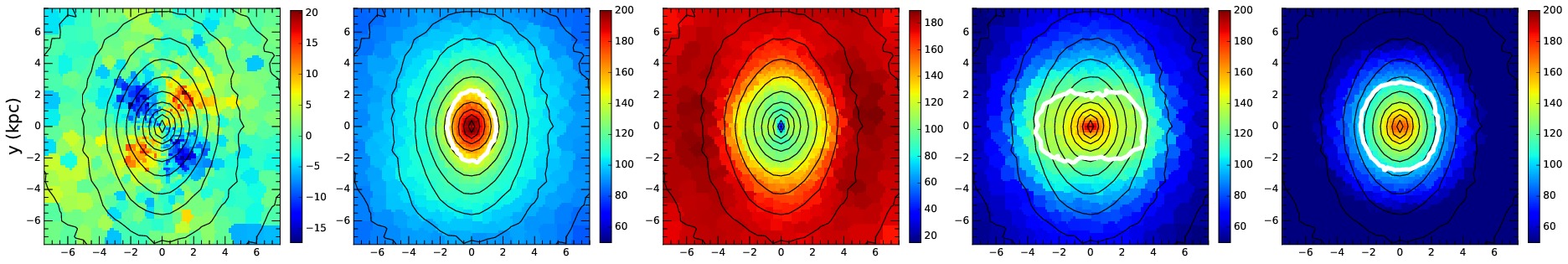}}
    \subfigure{\includegraphics[width=1.0\textwidth]{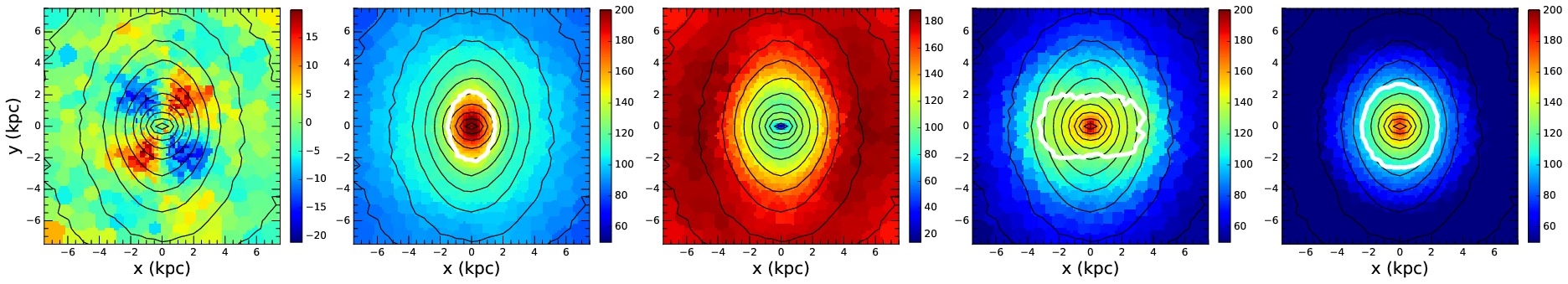}}
    \caption{The face-on view of the kinematics in cylindrical coordinates, from 
    left to right, 
    $\overline{v_R}, \sigma_R, \overline{v_\phi}, \sigma_\phi$, and $\sigma_z$. 
    The first row shows the SB model. The second and third rows show 
    the parallel and perpendicular case, respectively, of two bars in 
    the S2B model. The whole large-scale bars of S2B and SB models are covered. 
    Isodensity contours are overlaid in black. To highlight the 
    distribution of velocity dispersions, we also overlay their 
    $0.5(\sigma_\mathrm{max} + \sigma_\mathrm{min})$ contour with a 
    thick white line.}
    \label{fig:S2BcylOB}
    \subfigure{\includegraphics[width=1.0\textwidth]{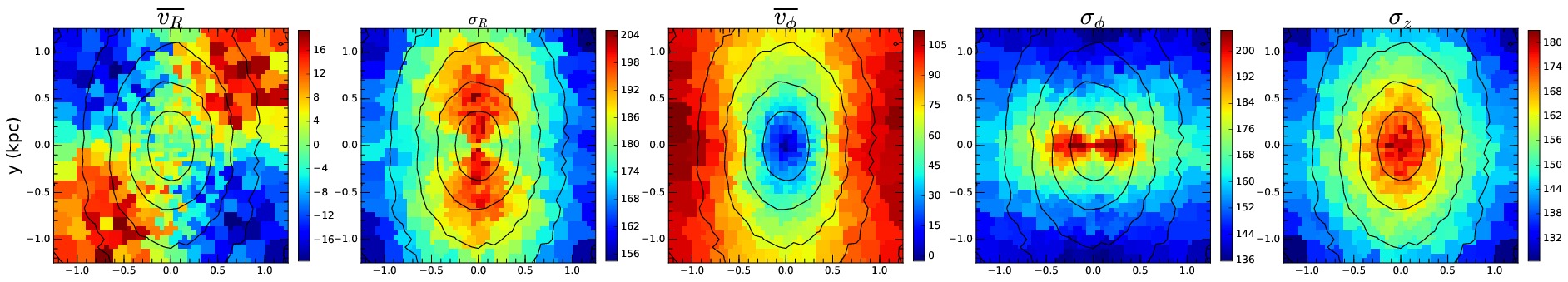}}
    \subfigure{\includegraphics[width=1.0\textwidth]{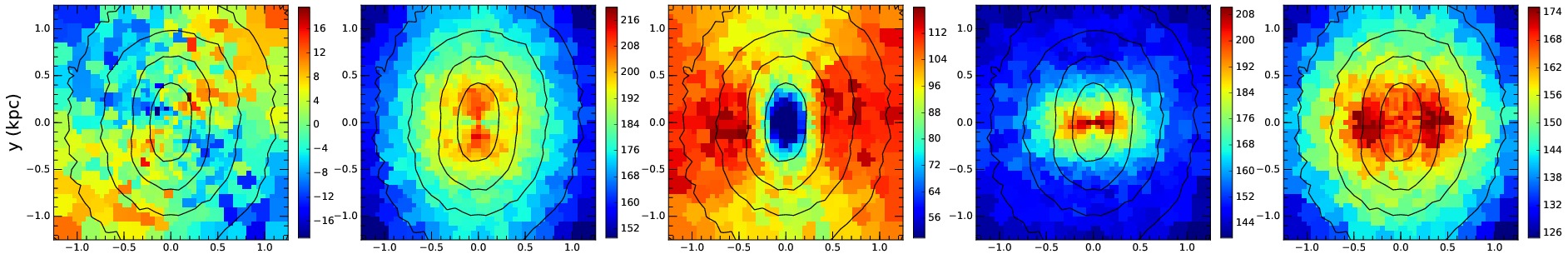}}
    \subfigure{\includegraphics[width=1.0\textwidth]{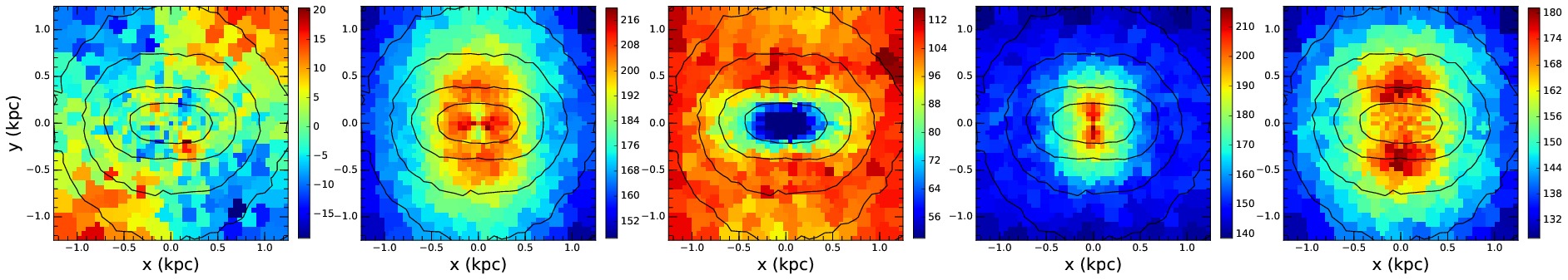}}
    \caption{Same as \reffig{fig:S2BcylOB}, zoomed into the central regions dominated by the inner bar.}
    \label{fig:S2BcylIB}
\end{figure*}

As shown above, S2B galaxies have very different nuclear kinematics
from single-barred galaxies. The inner bar in S2Bs exhibits many
distinguishable properties that have never been found in a large single
bar. Do these results suggest that the inner bar is
essentially a different structure to normal bars? Because of the
superposition of different components along the line of sight, the
structural information is tangled in complex LOSVDs. To improve our
understanding of S2Bs, we show the intrinsic kinematics of the outer
(\reffig{fig:S2BcylOB}) and inner (\reffig{fig:S2BcylIB}) bars in
the inertial frame, from left to right: $\overline{v_R}, \sigma_R,
\overline{v_\phi}, \sigma_\phi$, and $\sigma_z$. 

As we can see from \reffig{fig:S2BcylOB}, the large-scale bars
in the SB and S2B models have similar radial streaming motions,
i.e., butterfly-shaped radial motions ($\overline{v_R}$) extending 
to nearly half the length of the bars. As mentioned in \refsec{subsection:models},
the rotation curve of the models here flattens at $v_c\sim222$
km/s. It is clear that the stars in the outer disk rotate at 
$\sim v_c$, and the disk thus dominated by nearly circular motions. 
The large-scale bars in both the SB and S2B models extend close 
to its corotation radius. At the ends of the large-scale bars stars 
rotate at roughly $\sim v_c$. In the inner regions of the large-scale 
bars, the tangential velocity decreases significantly. Thus the random motions 
gradually become more important. 
We highlight the velocity dispersion contours at
$0.5(\sigma_\mathrm{max} + \sigma_\mathrm{min})$ with thick white 
curves. For large-scale bars, the radial velocity dispersion
$\sigma_R$ is always aligned with the bar; the tangential velocity
dispersion $\sigma_\phi$ is perpendicular to the bar. We
compare the case when the two bars are parallel (the middle row) 
with when they are perpendicular (the bottom
row). The relative orientation of the two bars has no significant
effect on the intrinsic motions of the outer bar in the S2B
model. Compared with the large-scale bar in the SB model (the top
row), the outer bar in the S2B model has no distinguishable
differences.

In \reffig{fig:S2BcylIB}, we zoom into the central regions of
\reffig{fig:S2BcylOB}. As expected, the inner bar generates
$\overline{v_R}$ and $\overline{v_\phi}$ features decoupled from the
outer bar (the middle and bottom rows), while the bar in the SB
model (the top row) acts as a solid body. For the inner bar in 
the S2B model, $\overline{v_\phi}$ shows significant local maxima 
along the minor axis. Such features clearly show that along the minor 
axis stars participate in the high-speed elongated motions, which is 
consistent with the expectation from the $h_3-\bar{v}$ 
anti-correlation discussed in \refsec{subsec:h3h4}. In spite of the 
differences in their sizes and pattern speeds, the inner bar in 
the S2B model and the large-scale bar in the SB model present similar 
$\sigma_R$ and $\sigma_\phi$ enhancements along the major and minor 
axis, respectively. From this point of view, inner bars have
qualitatively similar intrinsic motions as large-scale bars, being
essentially scaled-down versions of normal large-scale
bars. If that were the case, we would expect that the
inner bar exhibits similar $\sigma$ features as large-scale bars
and lacking $\sigma$-humps.
However, as shown in the rightmost column in \reffig{fig:S2BcylIB},
the inner bar presents $\sigma_z$-humps on its minor axis, while
$\sigma_z$ smoothly decreases outward in the SB model. Therefore, the
$\sigma$-humps seen in \reffig{fig:S2Bsigma} must be related to the
$\sigma_z$-humps, as they are the only difference with the SB model.

\begin{figure}[htp]
        \subfigure{\includegraphics[width=0.23\textwidth]{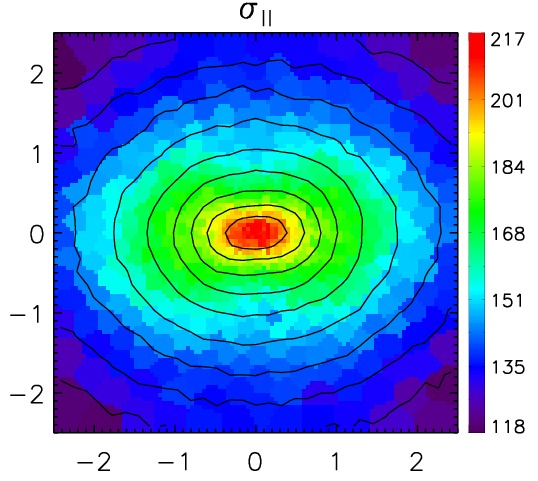}}
        \subfigure{\includegraphics[width=0.23\textwidth]{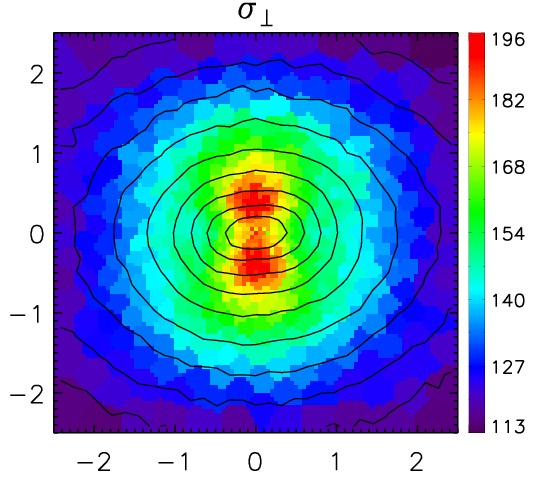}}
    \caption{The face-on view of velocity dispersions in the parallel ($\sigma_\parallel$) 
    and perpendicular ($\sigma_\perp$) directions with respect to the inner bar in the S2B 
    model. Both the outer and the inner bars are aligned with the $x$-axis.} 
    \label{fig:S2BCart}
\end{figure}

To better understand $\sigma$-humps appearing in S2Bs, because the bar is a
symmetric structure, it is more convenient to analyse the contributions 
of the parallel ($\sigma_\parallel$) and perpendicular
($\sigma_\perp$) components with respect to the inner bar. In
\reffig{fig:S2BCart} the inner bar is aligned with the $x$-axis;
thus $\sigma_\parallel$ and $\sigma_\perp$ are equal to $\sigma_x$ and
$\sigma_y$, respectively. As shown here, $\sigma_\perp$ humps are
present on the minor axis, while the elliptical $\sigma_\parallel$
peak is aligned with the bar. When the inner bar is parallel to
the LON, the line-of-sight velocity dispersion $\sigma_\mathrm{LOS}$
is mainly contributed by $\sigma_\perp$ and $\sigma_z$, by
$\sigma^2_\mathrm{LOS} = \sigma^2_\perp \mathrm{sin}^2i + \sigma^2_z
\mathrm{cos}^2i$. At small inclination, $\sigma_\mathrm{LOS}$ is
determined by $\sigma_z$, while $\sigma_\perp$ becomes more and more
important with increasing inclination. Because both $\sigma_\perp$
and $\sigma_z$ have significant humps at the minor axis,
$\sigma$-humps are present at any inclination when the inner bar
is parallel with the LON. When the inner bar is perpendicular to
the LON, the $\sigma_\mathrm{LOS}$ is contributed by $\sigma_\parallel$ 
and $\sigma_z$, by $\sigma^2_\mathrm{LOS} = \sigma^2_\parallel \mathrm{sin}^2i 
+ \sigma^2_z \mathrm{cos}^2i$. Because the $\sigma_\parallel$ enhancement
is aligned with the major axis of the inner bar, while $\sigma_z$ humps 
appear on the minor axis, the combination of $\sigma_\parallel$ and
$\sigma_z$ makes $\sigma_\mathrm{LOS}$ quite axisymmetric at intermediate 
inclinations. Thus $\sigma$-humps become barely visible for the case of 
$\mathrm{PA}_\mathrm{rel}=90^\circ, i=60^\circ$ shown in \reffig{fig:S2Bsigma}.

Therefore, the $\sigma_\perp$ humps and $\sigma_\parallel$ peak appearing 
on the minor and major axis, respectively, are normal kinematics of stellar bars. 
The properties of observed $\sigma_\mathrm{LOS}$-humps can be explained by the 
superimposition between $\sigma_z$-humps and such normal kinematics of bars. 
$\sigma_z$-humps play an important role in generating observable $\sigma$-humps. 
The physical origin of such $\sigma$ features will be studied in a follow-up 
study (Du et al., in preparation).

\section{The kinematic effect of a nuclear disk}
\label{section:discu}

As suggested by recent numerical simulations \citep{du_etal_15, woz_15}, the inner 
bar may form from 
the bar instability of a dynamically cold nuclear disk which forms from gas accumulation, 
followed by a starburst. This formation scenario indicates that a nuclear-disk-bar system 
where an inner bar is embedded in a rotation-dominated nuclear disk may be very common. 
In this paper, we show that the existence of a inner bar alone can explain the 
observed kinematics of S2Bs, especially $\sigma$-humps/hollows and positive $h_4$ rings. 
Our S2B model did not include the kinematic effect of a nuclear disk which 
is well known as a mechanism for reducing $\sigma$ 
\citep[e.g.][]{ems_etal_01, col_etal_14}. There is no doubt that a dynamically cooler 
nuclear disk, forming from the inflow of gas, can cause central $\sigma$ drops. Being 
embedded in a 
dynamically hotter bar or bulge, $\sigma$ is lower over the projected regions of the 
nuclear disk, while beyond the nuclear disk $\sigma$ is relatively larger. As shown in 
\reffig{fig:Atlas3Dno}, \object{NGC 3945} is a prototypical example of the kinematics of a strong 
nuclear disk. \citet{col_etal_14} compared this galaxy extensively with their $N$-body+gas 
simulation. 
Within the projected nuclear disk, the reduced $\sigma$ value in \object{NGC 3945} is quite flat
except for the very central regions, which is consistent with the $N$-body+gas simulation 
described in \citet{col_etal_14}. The relatively larger $\sigma$ on the minor axis may give 
an impression of ``$\sigma$-humps'' perpendicular to the LON. 

Without a nuclear disk, the kinematics of the S2B model are mainly dominated by the velocity 
dispersion. For \object{NGC 2859} and \object{NGC 2950}, the nuclear disk causes significant 
local minima and maxima in $\bar{v}$ fields and an $h_3-\bar{v}$ anti-correlation, while we 
cannot see any clear effects of the nuclear disk in the $\sigma$ maps. The $\sigma$ features can 
be well matched by the S2B model without a nuclear disk, which suggests that such 
nuclear disks may not efficiently reduce the $\sigma$ value in these S2Bs. The nuclear disk 
in \object{NGC 3384}, if it exists, seems even weaker than in \object{NGC 2950}, thus it 
may not be sufficient for forming $\sigma$-hollows (or $\sigma$-humps) as large as 30 km/s. 
Although we cannot rule out the scenario 
that $\sigma$-humps are caused by a nuclear disk, the S2B model does share similar kinematic 
features with \object{NGC 3384}. It is reasonable to consider \object{NGC 3384} 
as a potential S2B candidate. Further numerical simulations are required to better quantify the 
kinematic effect of a dynamically cold nuclear disk, especially for nuclear-disk-bar systems.

\section{Summary}
\label{section:conclusion}

This study sheds new light on the kinematic properties of
double-barred galaxies. Using well-resolved, self-consistent
simulations, we have studied the kinematic properties of double-barred
galaxies in comparison to single-barred galaxies. By quantifying the LOSVDs
with Gauss-Hermite moments, we find that many significant
kinematic features are closely associated with the inner bar. The
most notable feature is $\sigma$-humps that appear on the
minor axis of inner bars, matching well with the integral-field observations of
the stellar kinematics from the \atlas and \sauron \ surveys. Accompanied by 
$\sigma$-ring/spiral-like features, $\sigma$-humps may help to explain the ubiquitous 
$\sigma$-hollows in S2Bs seen in previous observations. 
Generally, 
$\sigma$-humps evolve and oscillate together with the inner bar. Based on 
the analysis of intrinsic motions of bars, we show that the inner 
bar is essentially a scale-down version of normal large-scale bars from the kinematic 
point of view. The only difference is the $\sigma_z$-humps appearing on the minor axis 
of the inner bar. Combined with $\sigma_\parallel$ enhancements and 
$\sigma_\perp$ humps produced in normal bars, $\sigma_z$-humps are the 
key to generating the observed $\sigma$-humps in S2Bs.

The isovelocity contours are significantly distorted. However, at the
central regions, the kinematic major axis is only slightly distorted
toward the opposite direction with respect to the inner bars. The
most significant asymmetric twists are present at
intermediate radii, in the transition region between the two bars
instead of the photometric end of the inner bar. Because of the
elongated streaming motions in bars, some non-Gaussian features
appear. The outer bar exhibits an $h_3-\bar{v}$ correlation, as
expected. However, in the central regions, $h_3$ becomes
anti-correlated with $\bar{v}$ as a result of the increasing dominance
of the inner bar. The inner bar exhibits significant positive 
$h_4$ rings in nearly face-on cases, suggesting that the inner bar has a 
sharply peaked $v_z$ distribution.

\begin{acknowledgments}
M.D. thanks the Jeremiah Horrocks Institute of the University of
Central Lancashire for their hospitality during a three month visit
while this paper was in progress. Hospitality at APCTP during the 7th
Korean Astrophysics Workshop is kindly acknowledged. We thank Peter
Erwin for constructive discussions and providing us the WIYN images of the target 
galaxies, Adriana de Lorenzo-C$\acute{a}$ceres and Martin Bureau for constructive 
comments and discussions on the manuscript. M.D. also warmly thanks 
Sarah Bird for help with the language of this paper.  
The research presented here is partially supported by the 973 Program of 
China under grant no. 2014CB845700, by the National Natural Science 
Foundation of China under grant nos.11333003, 11322326, and by the 
Strategic Priority Research Program ``The Emergence of Cosmological 
Structures'' (no. XDB09000000) of the Chinese Academy of Sciences. We 
acknowledges support from a {\it Newton Advanced Fellowship} awarded by 
the Royal Society and the Newton Fund. This work made use of the facilities 
of the Center for High Performance Computing at Shanghai Astronomical 
Observatory. V.P.D. is supported by STFC Consolidated grant \# ST/J001341/1.  
V.P.D. was also partially supported by the Chinese Academy of Sciences 
President's International Fellowship Initiative Grant (No. 2015VMB004).
M.C. acknowledges support from a Royal Society University 
Research Fellowship.
\end{acknowledgments}


\end{document}